\documentclass[acmtocs, authorversion, manuscript, screen]{acmart}
\usepackage{booktabs} % For formal tables

\usepackage[ruled]{algorithm2e} % For algorithms

\acmJournal{TOPS}
\acmVolume{1}
\acmNumber{1}
\acmArticle{1}
\acmYear{2019}
\acmMonth{1}

\setcopyright{acmcopyright}
\acmDOI{0000001.0000001}

\newcommand{\highlighttext}[1]{\textcolor{black}{#1}}
\newcommand{\highlighttable}{\color{black}}

\newcommand{\htext}[1]{\highlighttext{#1}}
\newcommand{\htable}{\highlighttable}

\newcommand{\hdifftext}[1]{\textcolor{black}{#1}}

\newcommand{\diffminor}[1]{\textcolor{black}{#1}}

\usepackage{pifont}% http://ctan.org/pkg/pifont
\newcommand{\cmark}{\ding{51}}%
\newcommand{\xmark}{\ding{55}}%

\usepackage{lmodern}

\usepackage{units}
\usepackage{siunitx}

\usepackage{amssymb}

\usepackage[T1]{fontenc}

\usepackage{version}
\includeversion{mynote}
\usepackage{comment}

\usepackage{setspace}

\usepackage{graphicx}
\usepackage{tabularx}
\usepackage{array,multirow}

\usepackage[noend]{algpseudocode}
\usepackage{varwidth}
\usepackage{amsmath}

\usepackage{subcaption}

\usepackage{pgffor}
\newcommand{\repeatvalue}[2]{% \repeat already defined
    \foreach \n in {1,...,#1}{#2}
}

\newcommand{\whitehspace}{\hspace{2mm}}

\usepackage{contour}
\contourlength{0.01em}% Smaller radius

\newcommand{\ANCHOR}{\textsc{anchor}\xspace}
\newcommand{\ANCHORBF}{\contour{black}{\textsc{Anchor}}\xspace}
\newcommand{\HOOKs}{\textsc{hook}s\xspace}
\newcommand{\Tamarin}{{\sc Tamarin}\xspace}

\newif\ifarxiv
\arxivfalse

\newtheorem{theo}{Theorem} %use: \begin{theo} \end{theo}

\newtheorem{lem}{Lemma}
\newtheorem{prop}{Proposition} %use: \begin{prop} \end{prop}
 
\newenvironment{proofm}{\noindent\emph{Proof:}}{\hfill$\Box$} 
\newcommand{\fourxspace}{\xspace\xspace\xspace\xspace}

\newcommand{\tablefontsize}{\scriptsize}

\begin{document}
% Title portion
\title{
ANCHOR: logically-centralized security for Software-Defined Networks
%%ANCHOR: a case of logically-centralized security for Software-Defined Networks
} 
% \titlenote{This is a titlenote}
% \subtitle{This is a subtitle}
% \subtitlenote{Subtitle note}

\author{Diego Kreutz}
\orcid{0000-0003-4491-646X}
\affiliation{%
  \institution{SnT, University of Luxembourg and Federal University of Pampa}
  \streetaddress{6, avenue de la Fonte}
  \city{Esch-sur-Alzette}
  \postcode{L-4364}
  \country{Brazil}
}
\author{Jiangshan Yu}
\affiliation{%
  \institution{SnT, University of Luxembourg and Monash University}
  \streetaddress{Wellington Rd}
  \city{Clayton VIC}
  \postcode{3800}
  \country{Australia}
}
\author{Fernando M. V. Ramos}
\affiliation{%
  \institution{LASIGE, Faculdade de Ci\^{e}ncias, Universidade de Lisboa}
  \city{Lisbon}
  \country{Portugal}
}
\author{Paulo Esteves-Verissimo}
\affiliation{%
  \institution{SnT, University of Luxembourg}
  \streetaddress{6, avenue de la Fonte}
  \city{Esch-sur-Alzette}
  \postcode{L-4364}
  \country{Luxembourg}
}

\renewcommand\shortauthors{Kreutz, D. et al}

\begin{abstract}

Software-defined networking (SDN) decouples the control and data planes of
traditional networks, logically centralizing the functional properties
of the network in the SDN controller. While this centralization
brought advantages such as a faster pace of innovation, it also
disrupted some of the natural defenses of traditional architectures
against different threats.
The literature on SDN has mostly been concerned with the functional side, despite
some specific works concerning non-functional properties like
`security' or `dependability'. Though addressing the latter in an
ad-hoc, piecemeal way, may work, it will most likely lead to
efficiency and effectiveness problems. 

We claim that the enforcement of non-functional properties as a pillar
of SDN robustness calls for a systemic approach.  We further advocate,
for its materialization, the re-iteration of the successful formula
behind SDN -- `logical centralization'.  As a general concept, we
propose \ANCHOR, a subsystem architecture that promotes the logical
centralization of non-functional properties. To show the
effectiveness of the concept, we focus on `security' in this paper: we
identify the current security gaps in SDNs and we populate the
architecture middleware with the appropriate security mechanisms, in a
global and consistent manner.
\htext{Essential security mechanisms provided by \ANCHOR include reliable entropy and resilient
  pseudo-random generators, and protocols for secure registration and
  association of SDN devices.}
  
We claim and justify in the paper that centralizing such mechanisms is
key for their effectiveness, by allowing us to: define and enforce
global policies for those properties; reduce the complexity of
controllers and forwarding devices; ensure higher levels of robustness
for critical services; foster interoperability of the non-functional
property enforcement mechanisms; \htext{and promote the
  security and resilience of the architecture itself.}  We discuss
design and implementation aspects, and we prove and evaluate our
algorithms and mechanisms, \htext{including the formalisation of the main protocols and the verification of their
  core security properties using the \Tamarin prover.}

\end{abstract}



\thanks{We would like to thank the anonymous reviewers for the insightful comments. 
This work is partially supported by the
Fonds National de la Recherche Luxembourg (FNR) through PEARL grant
FNR/P14/8149128, by European Commission 
funds through the H2020 programme, namely by funding of the SUPERCLOUD project, ref. H2020-643964 and by Funda\c{c}\~{a}o para a Ci\^{e}ncia e a Tecnologia (FCT - Portugal), namely by funding of LaSIGE Research Unit, ref. UID/CEC/00408/2013 and the uPVN project, ref. PTDC/CCI-INF/30340/2017.

Authors' addresses: D. Kreutz, P. Ver\'issimo, SnT, University
of Luxembourg, Campus Belval, 6, avenue de la Fonte, L-4364 Esch-sur-Alzette; email: kreutz@acm.org, paulo.verissimo@uni.lu; J. Yu,  Monash University, Wellington Rd, Clayton VIC 3800, Australia; email: Jiangshan.Yu@monash.edu; F. M. V. Ramos, LASIGE, Departamento de Inform\'atica, Faculdade de Ci\^{e}ncias, Universidade de Lisboa, 1749-016 Lisboa, Portugal; email: fvramos@ciencias.ulisboa.pt}

\maketitle

% Introduction
%%%%%%%%%%%%%%%%%%%%%%%%%%%%%%%%%%%%%%%%%%%%%%%%
\section{Introduction}
%%%%%%%%%%%%%%%%%%%%%%%%%%%%%%%%%%%%%%%%%%%%%%%%

\ifarxiv

% fast pace of innovation at the cost of heterogeneity
Software-defined networking (SDN) moves the control functions out of the forwarding devices, logically centralizing the functional properties of the network. 
This decoupling between control and data plane leads to higher flexibility and programmability of network control, enabling fast innovation.
The deployment of network applications as software artefacts that run on a logically centralized controller provides the agility of software evolution to networks, in contrast to the comparably slow innovation dictated by progress in hardware.
Moreover, as the forwarding devices are directly controlled by a centralized entity, control applications can reason based on a global network view, enabling improved network operation.
In spite of all these benefits, this decoupling, associated with a common southbound API (e.g., OpenFlow), has removed an important natural protection of traditional networks.
Specifically, the heterogeneity and diversity of configuration protocols, the closed (and proprietary) nature of the devices, and the distributed nature of the control plane.
For instance, an attack on traditional forwarding devices would need to compromise different protocol interfaces -- in SDN much harm can be done by attacking OpenFlow alone.
Hence, from a security perspective, SDN introduces new attack vectors and radically changes the threat surface~\cite{kreutz2013HotSDN,hayward2016asurvey,dacier2017sco}. 

\else

\hdifftext{
Software-defined networking (SDN) moves the control function out of the forwarding devices, logically centralizing the functional properties of the network. 
This decoupling between control and data plane leads to higher flexibility and programmability of network control, enabling fast innovation.
In spite of all these benefits, this decoupling, associated with a common southbound API (e.g., OpenFlow), has removed an important natural protection of traditional networks.
Specifically, the heterogeneity and diversity of configuration protocols, the closed (and proprietary) nature of the devices, and the distributed nature of the control plane.
Hence, from a security perspective, SDN introduces new attack vectors and radically changes the threat surface~\cite{kreutz2013HotSDN,hayward2016asurvey,dacier2017sco}. 
}

\fi

% non-functional properties
\diffminor{
So far, the SDN literature has been mostly concerned with \emph{functional} properties, such as improved routing and traffic engineering~\cite{Jain2013BEG,alvizu2017cst,lin2016control}, efficient topology discovery~\cite{pakzad2016efficient}, enhanced network security services~\cite{hu2014flowguard,qazi2013simple,scott2016survey,ahmad2015security}, among others, by exploiting the ability to program the control plane.
}

\diffminor{
\emph{Non-functional} properties are those related to reaching goals of quality of the operation of the global system, rather than to its specific behavior. However, SDN currently leaves the achievement of non-functional properties
to individual mechanisms or services. Works having recently seen the light, concerned in principle with non-functional properties, address specific implementations of \emph{functions} or \emph{services}, albeit
dependability-related~\cite{botelho2016design,ros2014five,Katta2015,kreutz2014sdnsurvey,berde2014onos} 
or
security-related~\cite{porras2012sek,shin2014rosemary,shin2013fresco,hayward2016asurvey}. 
To give an example, security services such as firewalls or Deep Packet Inspection (DPI) mechanisms for attack detection and mitigation, rely essentially on functional properties of the network, i.e., they are concerned with the SDN function of generating and remotely installing the appropriate flow rules in the data plane.
}

\diffminor{
%As effective as the former examples may be, their impact on the desired system-level non-functional property (say, integrity, or availability) ends-up being bottom-up, in an ad-hoc, piecemeal way. It may work, but most likely, it will create gaps in the enforcement of the property, which inevitably lead to efficiency and effectiveness problems (as we indeed show in Section~\ref{sec:background}).
As effective as the former examples may be, their impact on the desired system-level non-functional property (say, integrity, or availability) ends-up being bottom-up, in an ad-hoc, piecemeal way. 
It may work for specific cases, but generically, it is most likely to create gaps in the enforcement of the property, which would inevitably lead to efficiency and effectiveness problems (as we indeed show in 
Section~\ref{sec:background}).
For instance: insecure control plane associations or communications, network information disclosure, spoofing attacks, and hijacking of devices can easily compromise the network operation;
performance  crises can escalate to globally affect QoS;
unavailability and lack of reliability of controllers, forwarding devices, or clock synchronization parameters can considerably degrade network
operation~\cite{kloti2013openflow,akhunzada2015securing,hayward2016asurvey}.
%It is worth emphasizing that there are several security issues and challenges such as firewalls, deep packets inspection, attack detection and mitigation, software vulnerabilities on controllers and network apps (e.g. memory leak, lack of isolation between apps, implementation bugs, flow rule conflict resolution), and so forth~\cite{yoon2017flow,scott2016survey,ahmad2015security}. 
%Yet, software vulnerabilities are orthogonal to functional and non-functional properties.
}

\diffminor{
We claim that the coherent enforcement 
of non-functional properties is a pillar of SDN robustness, but it currently lacks a systemic, top-down approach.
As such, in this paper we propose a re-iteration of the successful formula behind SDN -- `logical centralization' -- for its materialization. We believe this step is critical to the successful deployment of SDN, especially at infrastructure/enterprise scale.
%In essence, \ANCHOR answers (only apparently redundant) questions, like: "How secure will security service ACME make my system?"
}

In fact, the problematic scenarios exemplified above can be best
avoided by the logical centralization of the system-wide enforcement
of non-functional properties, increasing the chances that the whole
architecture inherits them in a more balanced and coherent way.
The steps to achieve such goal are to: (a) select the crucial
properties to enforce (dependability, security, quality-of-service,
etc.); (b) identify the current gaps that stand in the way of achieving
such properties in SDNs; (c) design a logically-centralized subsystem
architecture and middleware, with hooks to the main SDN architectural
components, in a way that they can inherit the desired properties; \htext{and (d)
populate the middleware with the appropriate mechanisms and protocols
to enforce the desired properties/predicates, across controllers and
devices, in a global and consistent manner.}

% we propose logically-centralized non-functional property enforcement 

Generically speaking, it is worth emphasizing that centralization has
been proposed as a means to address different problems of current
networks.  For instance, the use of centralized cryptography schemes
and centralized sources of trust to authenticate and authorize known
entities has been pointed out as a solution for improving the security
of Ethernet networks~\cite{kiravuo2013survey}.  Similarly, recent
research has suggested network security as a service as a means to
provide the required security of enterprise
networks~\cite{hayward2016asurvey}.
However, centralization has its drawbacks, so let us explain why
centralization of non-functional property enforcement brings important
gains to software-defined networking. We claim, and justify ahead in
the paper, that it allows to define and enforce global policies for those properties, reduce the complexity of networking devices, ensure higher levels of robustness for critical services, foster interoperability of the non-functional enforcement mechanisms, and better promote the resilience of the architecture itself.

%%(e.g., registration, authentication and authorization)

%%%%%%%%%%%%%%%% SECURITY as use case %%%%%%%%%%%%%%%%%%%%%%

To achieve these goals, we propose \ANCHOR, a subsystem
architecture that does not modify the essence of the current SDN
architecture with its payload controllers and devices, but rather
stands aside,  `anchors' (logically-centralizes) crucial
functionality and properties, and `hooks' to the latter components, in
order to secure the desired properties.
The reader will note that this design philosophy concerns
any kind of non-functional properties. To prove our point, in this
paper we have chosen \textit{security} as our use case and identified
at least four gaps that stand in the way of achieving the former goals in
current SDN systems: (i) security-performance gap; (ii) complexity-robustness
gap; (iii) global security policies gap; and (iv) resilient
roots-of-trust gap.
The security-performance gap comes from the frequent conflict between
mechanisms enforcing those two
properties.
The complexity-robustness gap represents the conflict between the
current complexity of security and cryptography implementations, and the
negative impact this has on robustness and hence correctness.  The
lack of global security policies leads to ad-hoc and discretionary
solutions creating weak spots in architectures.  The lack of a
resilient root-of-trust burdens controllers and devices with trust
enforcement mechanisms that are ad-hoc, have limited reach and are
often sub-optimal.  We further elaborate in the paper on the reasons
behind these gaps, their negative effects in SDN architectures, and
how they can possibly be mitigated through a logically-centralized
security enforcement architecture. That is, in this particular case study,
the architecture middleware is populated with specific functionality
whose main aim is to ensure the `security' of control plane associations
and of communication amongst controllers and devices.

% \ANCHOR details
%%how they can possibly be mitigated through a logically-centralized
%%security enforcement subsystem architecture.

In addition, in this paper we give first steps in addressing a
long-standing problem, the fact that a single root-of trust --- like
\ANCHOR, but also like any other standard trusted-third-party, like
e.g., CAs in X.509 PKI or the KDC in \htext{Kerberos --- is a \emph{single
    point failure} (SPoF).}  There is nothing wrong with SPoFs, as
long as they fail rarely, and/or the consequences of failure can
be mitigated, which is unfortunately not the common case.  As such, we
start by carefully promoting reliability in the design of \ANCHOR,
endowing it with robust functions in the different modules, in order
to reduce the probability of failure/compromise.  Moreover, the
proposed architecture only requires symmetric key cryptography. This
not only ensures a very high performance, but also makes the system
secure against attacks by a quantum computer. Thus, the system is also
\emph{post-quantum secure}~\cite{bernstein2009ipqc}.  Second, we
mitigate the consequences of successful attacks, by protecting past,
pre-compromise communication, and ensuring the quasi-automatic
recovery of \ANCHOR after detection, even in the face of total control
by an adversary, achieving respectively, \emph{perfect forward secrecy
  (PFS)} and \emph{post-compromise security (PCS)}.  Third, \htext{since
  protocol designs are normally error prone, we formalise our protocol using a symbolic model 
  and verify its core properties using the \Tamarin~prover~\cite{Tamarin-cav}. Finally,} our architecture promotes resilience, or the continued
prevention of failure/compromise by automatic means. \htext{Though out of scope of this paper, the
  resilience of \ANCHOR using fault and
intrusion tolerance techniques is part of our plans for future work, as we
  discuss in Section~\ref{sec:rwork}.}

\ifarxiv

To summarize, the key contributions of our work include the following:
\begin{enumerate}

\item The concept of logical centralization of SDN non-functional
  properties provision.

\item The blueprint of an architectural framework based on middleware
  composed of a central `anchor', and local `hooks' in controllers
  and devices, hosting whatever functionality needed to enforce these properties.

\item A gap analysis concerning barriers in the achievement of
  non-functional properties in the security domain, as a
  proof-of-concept case study.

\item \htext{Definition, design and implementation of the mechanisms and
  algorithms to populate the middleware in order to fill those
  gaps, and achieve a logically-centralized
    security architecture that is reliable and efficient.}

\item \htext{The enforcement of strong properties such as post-quantum
    security, perfect forward secrecy, and post-compromise
    recovery. As we discuss in Section~\ref{sec:rwork}, these properties are not ensured by previous work and thus represent a clear advance over the state-of-the-art on SDN security.}

\item \htext{A formalisation of the main protocol, and a formal verification
    of its correctness and core security properties, through symbolic
    modelling using the \Tamarin prover.}

\item \htext{Evaluation of the proposed mechanisms.}

\end{enumerate}

\else

\hdifftext{
To summarize, the key contributions of our work include the following:
(1) the concept of logical centralization of SDN non-functional
  properties provision;
(2) the blueprint of an architectural framework based on middleware
  composed of a central `anchor', and local `hooks' in controllers
  and devices, hosting whatever functionality needed to enforce these properties;
(3) A gap analysis concerning barriers in the achievement of
  non-functional properties in the security domain, as a
  proof-of-concept case study;
(4) \htext{Definition, design and implementation of the mechanisms and
  algorithms to populate the middleware in order to fill those
  gaps, and achieve a logically-centralized
    security architecture that is reliable and efficient;}
(5) \htext{The enforcement of strong properties such as post-quantum
    security, perfect forward secrecy, and post-compromise
    recovery. As we discuss in Section~\ref{sec:rwork}, these properties are not ensured by previous work and thus represent a clear advance over the state-of-the-art on SDN security;}
(6) \htext{A formalisation of the main protocol, and a formal verification
    of its correctness and core security properties, through symbolic
    modelling using the \Tamarin prover;}
(7) \htext{Evaluation of the proposed mechanisms.}
}

\fi

We show that, compared to the state-of-the-art in SDN security, our
solution preserves at least the same security functionality, but
achieves higher levels of implementation robustness, by vulnerability
reduction, while providing high performance.  Whilst we try to prove
our point with security, our contribution is generic enough to inspire
further research concerning other non-functional properties (such as
dependability or quality-of-service).  
It is also worth emphasizing
that the architectural concept that we propose in this paper would
require a greater effort to be deployed in traditional networks, 
due to the heterogeneity of the infrastructure and its vertical 
integration. This will be made clear throughout the paper.
%since unlike what happens the SDN world,
%the former are rather heterogeneous, vertically integrated, complex
%and deprived of widely adopted open southbound
%APIs~\cite{kim2013inm,ramos2015sdn,kreutz2014sdnsurvey}, unlike e.g.,
%OpenFlow.  

%%Last, but not least, our major goal with \ANCHOR is to provide
%%security properties for control plane communications, i.e.,
%%associations and secure data exchange among controllers and
%%forwarding devices.
 
We have structured the paper as follows.  Section~\ref{sec:arch} gives
the rationale and presents the generic logically-centralized
architecture for the system-wide enforcement of non-functional
properties, and explains its benefits and limitations.  In
Section~\ref{sec:background}, we discuss the challenges and
requirements brought by the current gaps in security-related
non-functional properties.  Section~\ref{sec:arch-sec} describes the
logically-centralized security architecture that we propose, along
with its mechanisms and algorithms. \htext{The main algorithms are
  co-designed with a formal model, and the formal verification of their security properties is
  presented in Section~\ref{sec:security}.} Then, in
Sections~\ref{sec:implementation} and~\ref{sec:evaluation}, we discuss
design and implementation aspects of the architecture, and present
evaluation results.  In Sections~\ref{sec:rwork}
and~\ref{sec:discussion}, we give a brief overview of related work,
discuss some challenges and justify some design options of our
architecture.  Finally, in Section~\ref{sec:conclusion}, we conclude.

% Architecture
%%%%%%%%%%%%%%%%%%%%%%%%%%%%%%%%%%%%%%%%%%%%%%%%
\section{The \ANCHOR Architecture}
%%logically-centralized \\ non-functional properties

\label{sec:arch}
%%%%%%%%%%%%%%%%%%%%%%%%%%%%%%%%%%%%%%%%%%%%%%%%

\begin{figure}[ht]
   \centering
   \includegraphics[width=0.55\columnwidth]{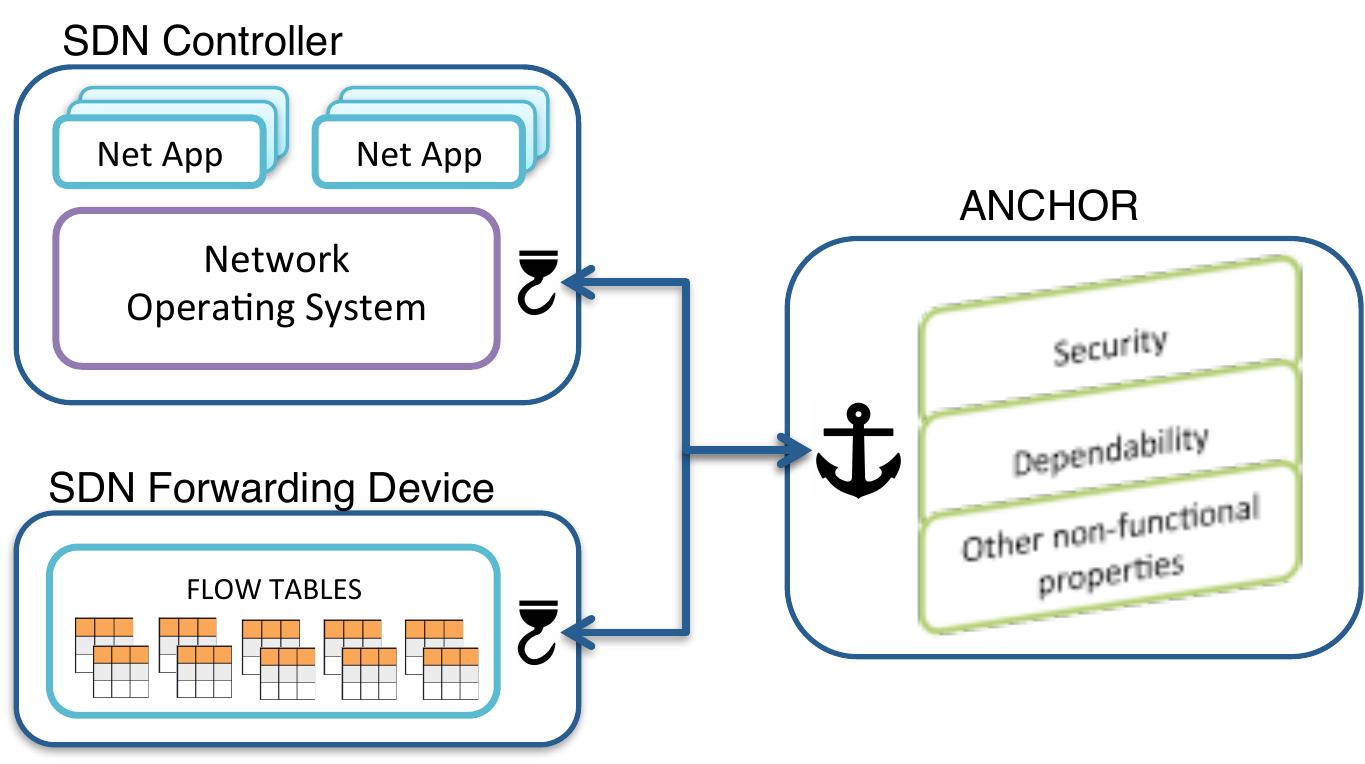}
   \caption{\ANCHOR general architecture}
   \label{fig:general}
\end{figure}

In this section we introduce \ANCHOR, a general architecture for
logically-centralized enforcement of non-functional properties -- such
as `security', `dependability', or `quality-of-service'
(Figure~\ref{fig:general}) -- in SDN.  The logical centralization of the
provision of non-functional properties allows us to: (1) define and
enforce global policies for those properties; (2) reduce the
complexity of controllers and forwarding devices; (3) ensure higher
levels of robustness for critical services; (4) foster
interoperability of the non-functional property enforcement
mechanisms; and finally (5) better promote the resilience of the
architecture itself.  Let us explain the rationale for these claims.

\textbf{Define and enforce global policies for non-functional properties.}  One
can enforce non-functional properties through piece-wise, partial
policies.  
\hdifftext{
But it is easier and less error-prone, as attested by SDN architectures with respect to the functional properties, to enforce
e.g., security policies, from a central trust point,
in a globally consistent way. Especially when one considers changing
policies during system lifetime. 
}

\textbf{Reduce the complexity of controllers and forwarding devices.}
\hdifftext{
One of the ideas of SDN was exactly to simplify
the construction of devices, by stripping them of functionality,
centralized on controllers.
We are extending the scope of the concept, by relieving both
controllers and devices from ad-hoc and redundant implementations of
mechanisms that are bound to have a critical impact on
the whole network.
}

\htext{
\textbf{Ensure higher levels of robustness for critical services.}
Enforcing non-functional properties like dependability or
security has a critical scope, as it potentially affects
the entire network. 
Unfortunately, the robustness of devices and controllers is still a concern, as 
they are becoming rather complex, which leads to several critical
vulnerabilities, as amply exemplified in~\cite{hayward2016asurvey}.  
%For these reasons, a single device or controller may become a single
%point of failure for the network.
A centralized concept as we advocate might considerably
improve on the situation, exactly because the enforcement of
non-functional properties would be
achieved through a specialized (carefully
designed and verified) subsystem, minimally interfering with
the SDN payload architecture.  
%A dedicated implementation, carefully
%designed and verified, would be re-usable, not re-implemented, by the
%payload components.
}

\htext{
\textbf{Foster interoperability of the non-functional property
  enforcement mechanisms.}  Different controllers require different
configurations today, and a potential lack of interoperability in
terms of non-functional properties arises. Having global policies and mechanisms for
non-functional property enforcement also creates an easier path to
foster controller and device interoperability (e.g., East and
Westbound APIs).  
%This way, mechanisms can
%be modified or added, and have a global repercussion, without the
%challenge of having to implement such services in each component.
}

\textbf{Better promote the resilience of the architecture itself.}
Having a specialized subsystem architecture already helps for a start,
since for example, its operation is not affected by latency and
throughput fluctuations of the (payload) control platforms
themselves. However, the considerable advantage of both the decoupling
and the centralization is that it becomes straightforward to design
in security and dependability measures for the architecture itself,
such as advanced techniques and mechanisms to tolerate
faults and intrusions (and in essence overcome the main disadvantage
of centralization, the potential single-point-of-failure risk).

\hdifftext{
The outline of our architecture is depicted in
Figure~\ref{fig:general}.
The ``logically-centralized'' perspective of non-functional property
enforcement is materialized through a subsystem architecture relying
on an anchor of trust, a middleware whose main aim
is to ensure that certain properties -- e.g., the security of
control plane associations and of communication amongst controllers
and devices -- are met throughout the architecture.
}

\htext{
In a manner similar to traditional security services, such as Kerberos and RADIUS, \ANCHOR is a set of services for the SDN architecture.
}  
It `anchors' crucial functionality and properties, and `hooks' to the
former components, in order to secure the desired properties.
So, on the devices, we just need the local counterparts to the \ANCHOR
middleware mechanisms and protocols, or \textsc{hook}s, to interpret and follow
the \ANCHOR's instructions.
\htext{In contrast to traditional services, however, \ANCHOR targets SDN infrastructures -- its advantage over existing systems is in part due to its specificity to this domain.} 

\hdifftext{
After having made the case for logically-centralized non-functional
property enforcement in SDN, and presenting the
outline of our general architecture, 
in the next two sections we introduce the use case we elected to show
in this paper, i.e., \textit{logically-centralized security}.  We start with a gap analysis that establishes the requirements for the
architecture functionality in Section~\ref{sec:background}, and then,
in Section ~\ref{sec:arch-sec}, we show how to populate \ANCHOR with the
necessary mechanisms and protocols to meet those requirements.
}

% Challenges

%%%%%%%%%%%%%%%%%%%%%%%%%%%%%%%%%%%%%%%%%%%%%%%%
\section{Challenges and requirements for security}
\label{sec:background}
%%%%%%%%%%%%%%%%%%%%%%%%%%%%%%%%%%%%%%%%%%%%%%%%

%To elaborate on our `security' case study, in
%this section we discuss, with more detail, the challenges brought in by
%the previously mentioned gaps --- (i) security-performance; (ii)
%complexity-robustness; (iii) global security policies; and (iv)
%resilient roots-of-trust --- as well as the requirements they put on a
%logically-centralized approach to enforcing security, as a
%non-functional system property.

%%%%%%%%%%%%%%%%%%%%%%%%%%%%%%%%%%%%%%%
\subsection{Security $vs$ performance}
%%%%%%%%%%%%%%%%%%%%%%%%%%%%%%%%%%%%%%%

% security-performance 
The security-performance gap comes from the conflict between ensuring
high performance and using secure primitives.  This gap affects directly the
control plane communication, which is the crucial link between
controllers and forwarding devices, allowing remote configuration of
the data plane at runtime.  Control channels need to provide high
performance (high throughput and low latency) while keeping the
communication secure.

\ifarxiv
\else

\hdifftext{
The latency experienced by control plane communication is particularly critical for SDN operation.
The increased latency is a problem \emph{per se}, in terms of reduced responsiveness, but may also limit control plane scalability, which can be particularly problematic in large datacenters~\cite{benson2010network}.
Most of the existing commercial switches already show low
control plane performance with TCP (e.g., a few hundred
flows/s~\cite{kreutz2014sdnsurvey}, Section V.A.).  Adding security
worsens the problem: previous works have demonstrated that the use of
cryptographic primitives has a perceivable impact on the latency of
sensitive communication in OpenFlow-enabled
networks~\cite{kreutz2016kiss,kreutz2017kiss}, in HTTPS
connections~\cite{naylor2014https}, among other examples.
}

\hdifftext{
Ideally, we would want both security robustness and performance on control plane
channels.
In the context of the
security-performance gap, some directions that we point to in our
architectural proposal ahead are, for instance: the careful selection
of cryptographic primitives~\cite{kreutz2016kiss}; the adoption of
cryptographic libraries exhibiting a good performance-security
tradeoff, such as NaCl~\cite{Bernstein2012TSI}; and/or mechanisms that 
allow cheap per-message one-time-key distribution~\cite{kreutz2016kiss,kreutz2017kiss}.
We return to these mechanisms later.
}

\fi

%%%%%%%%%%%%%%%%%%%%%%%%%%%%%%%%%%%%%%%
\subsection{Complexity $vs$ robustness}
\label{sec:nacl}
%%%%%%%%%%%%%%%%%%%%%%%%%%%%%%%%%%%%%%%

% complexity-robustness
\htext{
The complexity-robustness gap represents the conflict between the
current complexity of security based on cryptography and system implementations, and the
negative impact this has on robustness and hence correctness, hindering the ultimate goal. 
}

\ifarxiv

\htext{
In the past few years, several studies have recurrently shown critical misuse issues of cryptographic APIs of different TLS implementations~\cite{Egele2013ESC,buhov2015nsc,razaghpanah2017stu}.
One of the main root causes of these problems is the inherent complexity of traditional solutions and the knowledge required to use them without compromising security.
}
For instance, more than 80\% of the Android mobile applications present at least one instance of wrong use of the cryptographic APIs.
Recent studies have also found different vulnerabilities in TLS implementations 
and have shown that longstanding implementations, such as
OpenSSL\footnote{OpenSSL suffers from different fundamental issues such as too many legacy features accumulated over time, too many alternative modes as result of tradeoffs made in the standardization, and too much focus on the web and DNS names.},  including its extensive 
cryptography, \htext{are unlikely to be completely verified
in the near future~\cite{beurdouche2015messy,fan2016attacking}.}
To address this issue, a few projects, such as miTLS~\cite{bhargavan2013implementing} and 
Everest~\cite{bhargavan2017everest}, propose new 
and verified implementations of TLS. However, several challenges remain to be
addressed before having a solution ready for wide use~\cite{bhargavan2017everest}.

\else

\htext{
In the past few years, several studies have recurrently shown critical misuse issues of cryptographic APIs of different TLS implementations~\cite{Egele2013ESC,buhov2015nsc,razaghpanah2017stu}.
One of the main root causes of these problems is the inherent complexity of traditional solutions and the knowledge required to use them without compromising security.
}
\hdifftext{
For instance, recent reports have found different vulnerabilities in TLS implementations 
and have shown that long-standing implementations, such as
OpenSSL,
%\footnote{OpenSSL suffers from different fundamental issues such as too many legacy features accumulated over time, too many alternative modes as result of tradeoffs made in the standardization, and too much focus on the web and DNS names.}
including its extensive 
cryptography, \htext{is unlikely to be completely verified
in the near future~\cite{beurdouche2015messy,fan2016attacking}.}
To address this issue, a few projects, such as miTLS~\cite{bhargavan2013implementing} and 
Everest~\cite{bhargavan2017everest}, propose new 
and verified implementations of TLS. However, several challenges remain to be
addressed before having a solution ready for wide use~\cite{bhargavan2017everest}.
}

\fi

\htext{While the problem persists, the number of security incidents remains non-negligible.
Recent examples include vulnerabilities that allow the recovery of the secret key of OpenSSL at low cost~\cite{Yarom2014roe}, and timing attacks that explore vulnerabilities in both PolarSSL and
OpenSSL~\cite{Arnaud2013taa,brumley2011sca}.}
On the other hand, failures in classical PKI-based authentication and
authorisation subsystems have been persistently
happening~\cite{cromwell2017massive,pwc2014ucc,hill2013failures}, with the sheer complexity of those systems being considered one of
the root causes behind these problems.

\htext{
Similarly to the cryptographic APIs example, the leading cause of most security issues in systems -- and this includes SDN controllers, operating systems, hypervisors, etc. -- is the inherent complexity of their implementation and the amount of aggregated functions and services, resulting in challenging ecosystems in terms of security~\cite{dacier2017security,klein2009sel4,steinberg2010nova,ponemon2018the,arbettu2016security,yoon2017flow,secci2017onos,lee2017delta,singaravelu2006reducing,ho2003the}.
It is recognized by the community that complexity reduction (e.g., by means of isolation, modularization, reduced and verifiable code bases, loosely coupled and well-defined micro-services) plays a vital role in ensuring the security of systems.
}

\ifarxiv

\htext{
Considering the widely acknowledged principle that simplicity is key to
robustness, especially for secure systems, we advocate and try to
demonstrate in this paper, that the complexity-robustness gap can be
significantly reduced through less complex
but equally secure alternative solutions.
NaCl~\cite{Bernstein2012TSI}, which we mentioned in the previous section, can be used again as an example in this context: it is one of the first attempts to provide a less complex, efficient, yet secure alternative to OpenSSL-like implementations.
Mechanisms
simplifying key distribution, authentication and authorization, such
as iDVVs~\cite{kreutz2016kiss}, could help mitigate PKIs' problems.
Furthermore, simple and efficient protocols for ensuring the secure registration and association of devices are two other examples of reduced complexity when compared to traditional solutions (e.g., PKI/X.509 and TLS).
By following this direction, we are applying the same principle of
vulnerability reduction used in other systems, such as unikernels, where
the idea is to reduce the attack surface by generating a smaller
overall footprint of the operating system and
applications~\cite{williams2016unikernel}.
}

\else

\hdifftext{
Considering the widely acknowledged principle that simplicity is key to
robustness, especially for secure systems, we advocate and try to
demonstrate in this paper, that the complexity-robustness gap can be
reduced through less complex
but equally secure alternative solutions.
For instance, we will show that it is possible to reduce complexity (when compared to traditional solutions, such as PKI/X.509 and TLS), and yet \emph{improve} on security, by designing simple and efficient protocols for the secure registration and association of network devices.
By following this direction, we are applying the same principle of
vulnerability reduction used in other systems, such as unikernels, where
the idea is to reduce the attack surface by generating a smaller
overall footprint of the operating system and
applications~\cite{williams2016unikernel}.
}

\fi

%%%%%%%%%%%%%%%%%%%%%%%%%%%%%%%%%%%%%%%%%%%%%%%%
\subsection{Global security policies}
\label{sec:secreq}
%%%%%%%%%%%%%%%%%%%%%%%%%%%%%%%%%%%%%%%%%%%%%%%%

% policy 

The impact of the lack of global security policies can be illustrated
with different examples.  Although ONF describes data authenticity,
confidentiality, integrity, and freshness as fundamental requirements
to ensure the security of control plane communication in SDN, it does so in
an abstract way, and these measures are often ignored, or implemented
in an ad-hoc manner~\cite{hayward2016asurvey}. 
Another example is the lack of strong authentication
and authorisation in the control plane.  Recent reports show that
widely used controllers, such as Floodlight and OpenDaylight, employ
weak network authentication
mechanisms~\cite{wan2017arXivAFC,hayward2016asurvey,secci2017onos,lee2017delta}.
This leads to any forwarding device being able to connect to any
controller.  
%However, fake or hostile controllers or forwarding
%devices should not be allowed to become part of the network, in order
%to keep the network in healthy operation.  

From a security perspective, it is non-controversial that device identification, authentication
and authorization should be among the forefront requirements of any network.
All data plane devices should be appropriately
registered and authenticated within the network domain, with each
association request between any two devices (e.g., between a switch and a controller) being strictly
authorized by a security policy enforcement point.
In addition, control traffic should be secured, since it is the
fundamental vehicle for network control programmability.
This begs the question: why aren't these mechanisms employed in most deployments? 

\htext{
A reason for the current state of affairs is the lack of awareness, guidance, and enforcement policies.
It is therefore becoming crucial to define and establish global policies, and design, or adopt, the mechanisms needed to enforce them and meet the essential requirements (e.g., secure authentication and trustworthy authorization), to fill the policy gap.
%With policies put in place, it becomes easier to manage all network elements, concerning registration, authentication, authorization, and secure communication.
}

%%%%%%%%%%%%%%%%%%%%%%%%%%%%%%%%%%%%%%%
\subsection{Resilient roots-of-trust}
%%%%%%%%%%%%%%%%%%%%%%%%%%%%%%%%%%%%%%%

% root-of-trust

A globally recognized, resilient root-of-trust, could dramatically improve the global security of SDN, since current approaches to
achieve trust are ad-hoc and partial~\cite{abdullaziz2016light}.
\hdifftext{Solving this gap would certainly assist in fostering global mechanisms to ensure trustworthy registration and association between devices, as discussed before, but the benefits would go beyond that.}
For instance, a root-of-trust can be used to provide fundamental mechanisms (e.g., sources of
strong entropy or pseudo-random generators (PRGs)), which would serve as building blocks for specific security functions.

As a first example, modern cryptography relies heavily on strong keys
and the ability to keep them secret.  The core feature that defines a
strong key is its randomness.  However, the randomness of keys is
still a widely neglected 
issue~\cite{Vassilev2014tie} and, not surprisingly, weak entropy, 
and weak random number generation have 
been the cause of several significant vulnerabilities~\cite{Kim2013PAO}.  
Even long-standing cryptographic libraries such as OpenSSL have been
recurrently affected by this
problem~\cite{Kim2013PAO,openssl20161110}. 
Importantly, recent research has shown that this problem also affects networking equipment~\cite{Heninger2012MYP,albrecht2015factoring,hastings2016weak}.
For instance, a common pattern found in low-resource devices, such as switches, is that 
the random number generator of the operating system may lack the input of external 
sources of entropy to generate reliable cryptographic keys.

\ifarxiv

As a second example, sources of accurate time, such as the
local clock and the network time protocol, have to be secured to avoid
attacks that can compromise network operation, since time
manipulation attacks (e.g., NTP attacks~\cite{malhotra2015attacking,stenn2015snt})
can affect the operation of controllers and applications.  For
instance, a controller can be led to deliberately disconnect
forwarding devices if it wrongly perceives the expiration of heartbeat
message timeouts.

\fi

\hdifftext{
It is worth emphasizing that the resilient roots-of-trust gap lies
exactly in the relative trust that can be put in partial
ad-hoc implementations of critical functions by controller developers
and manufacturers of devices, in contrast to a careful,
once-and-for-all architectural approach that can be reinstantiated in
different SDN deployments. The list not being exhaustive, we claim
that strong sources of entropy, resilient,
indistinguishable-from-random number generators, and accurate,
non-forgeable global time services, are fitting examples of such
critical functions to be provided by logically-centralized
roots-of-trust, helping close the former gap.
}

% Security, yet performance
% \input{lysbon_03_security_reqs.txt}

% Security Architecture

%%%%%%%%%%%%%%%%%%%%%%%%%%%%%%
\section{Logically-centralized security}
\label{sec:aot}\label{sec:arch-sec}
%%%%%%%%%%%%%%%%%%%%%%%%%%%%%%

\begin{figure}[ht]
   \centering
   \includegraphics[width=0.65\columnwidth]{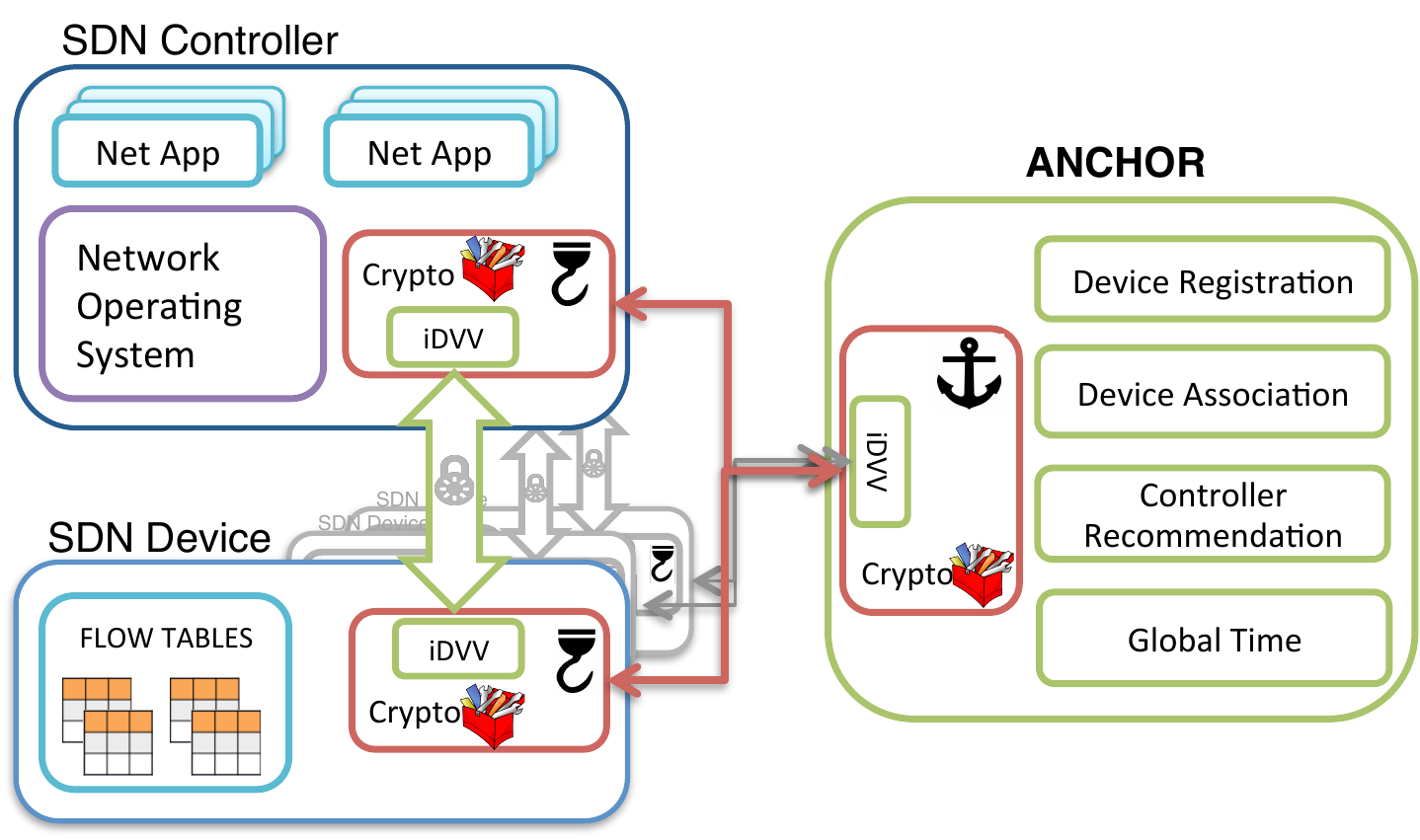}
   \caption{Logically-centralized Security}
   \label{fig:architecture}
\end{figure}

In this section we introduce the specialization of the \ANCHOR architecture for
logically-centralized security properties enforcement
(Figure~\ref{fig:architecture}), guided by the conclusions from the
previous section.
Our main goal is to provide security properties such as authenticity,
integrity, and confidentiality for control plane communication.  To
achieve this goal, the \ANCHOR provides mechanisms (e.g., registration, 
authentication, a source of strong entropy, a PRG) required to fulfill some of the major security requirements of
SDNs.

As illustrated in Figure~\ref{fig:architecture}, we ``anchor'' the
enforcement of security properties on \ANCHOR, which provides
all the necessary mechanisms and protocols to achieve the goal.  It is
also a central point for enforcing security policies by means of
services such as device registration, device association, controller
recommendation, or global time, thereby reducing the burden on
controllers and forwarding devices, which just need the local
\HOOKs, protocol elements that interpret and follow the
\ANCHOR's instructions.

%%%%%%%%%%%%%%%%%%%%%%%%%%%%%

%%%%%%%%%% MECHANISMS

Next, we review the components and essential security services
provided by \ANCHOR. 
We first illustrate, in Section \ref{sec:Hardening_Anchor}, how we
implement our strategy of improving the robustness of \ANCHOR as a
single root-of-trust, by hardening \ANCHOR in the face of failures.
For example, concerning the mitigation of possible (though expectedly infrequent) security failures, we provide countermeasures such as Perfect Forward Secrecy (PFS) and Post-Compromise Security (PCS), protecting pre- and post-compromise communications in the presence of successful attacks.
Next, we propose a source of strong entropy (Section~\ref{sec:entropy}) and
a resilient pseudo random generator (Section~\ref{sec:prg}) for
generating security-sensitive materials.  These are crucial
components, as attested by the impact of vulnerabilities discovered in
the recent past, in sub-optimal implementations of the former in
several software
packages~\cite{bernstein2016dual,mimoso2016gpg,zetter2015backdoor,schneier2012lousy}. We
%packages~\cite{bernstein2016dual,mimoso2016gpg,zetter2015backdoor,nist2014removes,schneier2012lousy,argyros2012prng,ytosa2011system,mccullough2008microsoft,bello2008dsa}. We
implement and evaluate the robustness of these mechanisms.  \htext{We
  also leverage a recently proposed mechanism, the integrated device
  verification value (iDVV), to simplify authentication, authorization, and key generation amongst SDN components~\cite{kreutz2016kiss},
  which we review and put in the context of \ANCHOR
  (Section~\ref{sec:iDVV}).  }  Namely, the iDVV protocol runs between
the \ANCHOR, and the \HOOKs in controllers and switching devices. We
implement and evaluate iDVV generators for OpenFlow-enabled control
plane communication.  After defining system roles and its setup in
Section~\ref{sec:roles}, we present two essential services for secure
network operation --- device registration (Section~\ref{sec:devreg})
and device association (Section~\ref{sec:devassoc}) --- and we
describe how the above mechanisms interplay with our secure
device-to-device communication approach (Section~\ref{sec:devcom}).
%Then, we briefly
%introduce a secure global time NTP service
%(Section~\ref{sec:impl:time}), as another example of root-of-trust
%like services to be hosted in \ANCHOR.
\htext{
The list of services of \ANCHOR is certainly not closed.
One can think of other functionalities, such as tracking of forwarding devices association, alert generation in case of anomalous behaviours (e.g., recurrent reconnections), and so forth.}

%These ancillary management tasks are essential to keep track of the network operation status.
\htext{In what follows, we describe the main \ANCHOR services in detail.
To help the reader following our descriptions, we summarize the most important notations used in Table~\ref{tabNotations}.}

{\renewcommand{\arraystretch}{1.4}
\newcommand{\wsRT}{\repeatvalue{1}{\whitehspace}}
\begin{table*}[!htp]
\caption{Summary of notations}
\label{tabNotations}
\highlighttable
\begin{center}
\small
\begin{tabularx}{0.80\linewidth}{>{\tablefontsize}p{1.4cm}>{\tablefontsize}X>{\tablefontsize}p{3cm}}
\hline

\textbf{Notation} 	& \textbf{Description} 								& \textbf{Example} \\\hline

$H$       			& Cryptographic hash function 							& SHA512 \\\hline
$MAC$     		& Message Authentication Code algorithm 				& Poly1305 \\\hline
$X, Y$			& One entity belonging to \{$A$, $Di$, $M$, $C$, $F$\}		& Device (e.g., switch) $i$ \\\hline
$Ke_{XY}$		& Encryption secret key used between entities X and Y		& 256 bits random key \\\hline
$Kh_{XY}$		& MAC/HMAC secret key used between entities X and Y		& 256 bits random key \\\hline
$E_{XY}$ 			& Encryption primitive using secret key $Ke_{XY}$			& AES \\\hline
[],$HMAC_{XY}$ 	& keyed-Hash MAC of message [] using secret key $Kh_{XY}$	& HMAC-SHA512 \\\hline
KDF			 	& Key Derivation Function								& OpenSSL PBKDF2 \\\hline

\end{tabularx}
\end{center}
\end{table*}
}

\subsection{Hardening ANCHOR}
\label{sec:Hardening_Anchor}

The compromise of a root-of-trust is of great concern, since crucial
services normally depend on it being secure and dependable.  
\hdifftext{
As we stated before, we have a long-term strategy towards the resilience of \ANCHOR.
In the context of this paper, it starts by improving the inherent reliability of its simplex (non-replicated) version, by hardening it in the face of failures.
For instance, different from existing traditional security services such as Kerberos and RADIUS, we still provide some security guarantees even when \ANCHOR has been compromised. 
}
In particular, we propose protocols to achieve two
security properties guaranteeing respectively, the security of past
(pre-compromise) communications, and of future (post-recovery)
communications. 
This provides a significant improvement over other
existing root-of-trust infrastructures.

The first security property is \emph{perfect forward secrecy (PFS)},
namely, the assurance that the compromise of all secrets in a current
session does not compromise the confidentiality of the communications
of the past sessions. The enforcement of PFS is systematically
approached in the algorithms we present next.

The second property is \emph{post-compromise security (PCS)}. While
PFS considers how to protect the past communications, PCS considers
how to automatically reinstate and re-establish the secure
communication channels, for future communications. This security
property has so far been considered only in the specific scenario of
secure messaging \cite{YuR-attacker-model-15}, and only limited works
\cite{DECIM-report15,YU2016DECIM} are available. In particular, we
consider that when \ANCHOR has been compromised by an attacker
(e.g., through the exploitation of software vulnerabilities), and has
been reinstated by the operator (e.g., by applying software patches and
rebuilding servers), the system should have a way to automatically
re-establish secure communications between \ANCHOR and all other
participants, without having to reinstate these components
(controllers and forwarding devices in this case, whose shared secrets
became compromised).
In particular, in Section \ref{sec:post_compromise} we explain how to
re-establish secure communication channels in a semi-automatic way,
after complete failure of \ANCHOR.

\iftrue

In summary, even though \ANCHOR is a single root-of-trust in our
system, we mitigate the associated risks by guaranteeing:

\begin{itemize}
\item PFS: the compromise of \ANCHOR in the current session does not
  expose \emph{past communications};
\item PCS: when \ANCHOR is compromised and reinstated, \ANCHOR can
  automatically re-establish secure communication channels with all
  other participants in the system to protect the security of
  \emph{future communications}.
\end{itemize}

\else

In summary, even though \ANCHOR is a single root-of-trust in our
system, we mitigate the associated risks by guaranteeing:
(a) PFS: the compromise of \ANCHOR in the current session does not
  expose \emph{past communications};
(b) PCS: when \ANCHOR is compromised and reinstated, \ANCHOR can
  automatically re-establish secure communication channels with all
  other participants in the system to protect the security of
  \emph{future communications}.

\fi

As a side note, since our system only uses symmetric key cryptography,
it will stand up even against an attacker with quantum computers. In
other words, our infrastructure will be \emph{post-quantum secure (PQS)}.

%%%%%%%%%%%%%%%%%%%%%%%%%%%%%%
\subsection{A source of strong entropy}
\label{sec:entropy}
%%%%%%%%%%%%%%%%%%%%%%%%%%%%%%

\hdifftext{
Entropy still represents a challenge for modern computers because they have been designed to behave deterministically~\cite{Vassilev2014tie}.
Sources of true randomness (e.g., physical phenomena such as atmospheric noise) can be difficult to use because they work differently from a typical computer.
}

%Weak entropy is still one of the most common root causes of security incidents involving keys of poor quality on computing systems and particularly on networking devices~\cite{Vassilev2014tie,Kim2013PAO}.
%Protocols such as SSH, TLS, and IKE are significantly affected by weak entropy~\cite{CorriganGibbs2013EHR}.

\htext{
To avoid the pitfalls of weak sources of entropy, in particular in
networking devices, \ANCHOR provides a source of strong entropy to
ensure the randomness required to generate seeds, pseudorandom values,
secrets, among other cryptographic material. The strong source of entropy
%, implemented by the Algorithm discussed in the Appendix A of ~\cite{kreutz2017anchor}, 
has the following property:
}

%%%%%%%%%%%%%%
\hdifftext{
\textbf{Strong Entropy -} Every value \textit{entropy} returned by 
\textit{entropy\_get} is indistinguishable-from-random.
}

Algorithm~\ref{alg:entropyupdate} shows how the external (from other
devices) and internal (from the local operating system) sources of
entropy are kept updated and used to generate random bytes per
function call (\textit{entropy\_get()}).  The state of the internal
and external entropy is initially set by calling the
\textit{entropy\_setup(data)}.  This function requires an input data,
which can be a combination of current system time, process number,
bytes from special devices, among other things, and random bytes
(\textit{rand\_bytes()}) from a local (deterministic) source of
entropy (e.g., \textit{/dev/urandom}) to initialize the state of the
entropy generator.
As we cannot assume anything regarding the predictability of the input data, we use it in conjunction with a \textit{rand\_bytes()} function call (line 2).
A call to \textit{rand\_bytes()} is assumed to return (by default) 64 bytes of random data.

\iftrue

\begin{figure*}[ht]
  %\centering
  \begin{minipage}[l]{.55\textwidth}
    \begin{algorithm}[H]
	\caption{Source of strong entropy}
	\label{alg:entropyupdate}

	\begin{algorithmic}[1]
	\small
	% min of two sources (e.g., controller & switch)
	\State \texttt{entropy\_setup(data)}
		\State \fourxspace \texttt{e\_entropy $\gets$ rand\_bytes()  $\oplus$ H(data)}
		\State \fourxspace \texttt{i\_entropy $\gets$ rand\_bytes()  $\oplus$ e\_entropy}
	\vspace{2mm}
	\State \texttt{entropy\_update()}
		\State \fourxspace \texttt{e\_entropy $\gets$ H($P_i || P_j$) $\oplus$ i\_entropy}
		\State \fourxspace \texttt{E\_counter $\gets$ 0}
	\vspace{2mm}
	\State \texttt{entropy\_get()}
		\State \fourxspace \texttt{if E\_counter >= MAX\_LONG call entropy\_update()}
		\State \fourxspace \texttt{i\_entropy $\gets$ H(rand\_bytes() $||$ E\_counter)}
		\State \fourxspace \texttt{\textbf{entropy} $\gets$ e\_entropy $\oplus$  i\_entropy}
	\end{algorithmic}
  \end{algorithm}
  \end{minipage}
  
\end{figure*}

\else

\begin{tabular}{ll}
\htable
\parbox{0.50\textwidth}{
\begin{minipage}[l]{.50\textwidth}
    \begin{algorithm}[H]
	\caption{{\footnotesize Source of strong entropy}}
	\label{alg:entropyupdate}

	\begin{algorithmic}[1]
	\footnotesize
	% min of two sources (e.g., controller & switch)
	\State \texttt{entropy\_setup(data)}
		\State \fourxspace \texttt{e\_entropy $\gets$ rand\_bytes()  $\oplus$ H(data)}
		\State \fourxspace \texttt{i\_entropy $\gets$ rand\_bytes()  $\oplus$ e\_entropy}
	\vspace{1mm}
	\State \texttt{entropy\_update()}
		\State \fourxspace \texttt{e\_entropy $\gets$ H($P_i || P_j$) $\oplus$ i\_entropy}
		\State \fourxspace \texttt{E\_counter $\gets$ 0}
	\vspace{1mm}
	\State \texttt{entropy\_get()}
		\State \fourxspace \texttt{if E\_counter >= MAX\_LONG call entropy\_update()}
		\State \fourxspace \texttt{i\_entropy $\gets$ H(rand\_bytes() $||$ E\_counter)}
		\State \fourxspace \texttt{\textbf{entropy} $\gets$ e\_entropy $\oplus$  i\_entropy}
	\end{algorithmic}
  \end{algorithm}
  \end{minipage}
}
&
\htable
\parbox{0.5\textwidth}{
 \begin{minipage}[r]{.4\textwidth}
  \begin{algorithm}[H]
	\caption{{\footnotesize Pseudo-random generator}}
	\label{alg:PRG}
	\begin{algorithmic}[1]
	\footnotesize
	\State \texttt{PRG\_setup()}
		\State \fourxspace \texttt{seed $\gets$ entropy\_get()}
		\State \fourxspace \texttt{counter $\gets$ long\_uint(entropy\_get())}
		\State \fourxspace \texttt{call entropy\_update()}
		\State \fourxspace \texttt{nprd $\gets$ seed $\oplus$ entropy\_get()}
	\vspace{1mm}
	\State \texttt{PRG\_refresh()}
		\State \fourxspace \texttt{seed $\gets$ entropy\_get()}
		\State \fourxspace \texttt{counter $\gets$ long\_uint(entropy\_get())}
		\State \fourxspace \texttt{nprd $\gets$ H(seed $\Vert$ nprd $\Vert$ counter)}
	\vspace{1mm}
	\State \texttt{PRG\_next()}
		\State \fourxspace \texttt{counter $\gets$ counter - 1}
		\State \fourxspace \texttt{if counter <= 0 call PRG\_refresh()}
		\State \fourxspace \texttt{\textbf{nprd}  $\gets$ HMAC(seed, nprd $\Vert$ counter)}
	\end{algorithmic}
  \end{algorithm}
  \end{minipage}
}
\end{tabular}

\fi

Function \textit{entropy\_update()} uses as input the statistics of external sources and the \ANCHOR's own packet arrival rate to update the external entropy.
The noise (events) of the external sources of entropy is stored in 32 pools ($P_0$, $P_1$, $P_2$, $P_3$, ..., $P_{31}$), as suggested by previous work~\cite{ferguson2011cryptography}.
Each pool has an event counter, which is reset to zero once the pool is used to update the external entropy. 
At every update, two different pools of noise ($P_i$ and $P_j$) are used as input of a hashing function $H$.
The two pools of noise can be randomly selected, for instance.
The output of this function is XORed with the internal entropy to generate the new state of the external entropy.
It is worth emphasizing that  \textit{entropy\_update()} is automatically called when \textit{E\_counter} (the global event counter) reaches its maximum value and whenever needed, i.e., the user can define when to do the function call. 

The resulting 64 bytes of entropy, indistinguishable-from-random bytes
(\textit{entropy\_get()}), are the outcome of an XOR operation between
the external and internal entropy.  While the external entropy
provides the unpredictability required by strong entropy, the internal
source provides a good, yet predictable~\cite{Vassilev2014tie},
continuous source
of entropy.  At each time the \textit{entropy\_get()} function is
called, the internal entropy is updated by using a local source of
random data, which is typically provided by a library or by the
operating system itself, and the global number of events currently in
the 32 pools of noise ($E\_counter$).  These two values are used as
input of a hashing function $H$.
%Thus, an attacker will have to compromise both the external and
%internal sources of entropy to compromise the resulting entropy.

Such sources of strong entropy can be achieved in practice by combining different
sources of noise, such as the unpredictability of network
traffic~\cite{Greenberg2009VSF}, the unpredictability of idleness of
links~\cite{Benson2010UDC}, packet arrival rate of network
controllers, and sources of entropy provided by operating systems. We
provide implementation details in Section~\ref{sec:strong:entropy}.
%
%%%%%%%%%%%%%%%%%%%%%%% PROOFS %%%%%%%%%%%%%%%%%%
%
A discussion about the correctness of Algorithm~\ref{alg:entropyupdate} can be found in Appendix A of ~\cite{kreutz2017anchor}. %~\ref{appendix:entropyupdate}
%$patime = current arrival time XOR last arrival time$

%%%%%%%%%%%%%%%%%%%%%%%%%%%%%%
\subsection{Pseudorandom generator (PRG)}
%%%%%%%%%%%%%%%%%%%%%%%%%%%%%%
\label{sec:prg}

A source of entropy is necessary but not sufficient. Most
cryptographic algorithms are highly vulnerable 
 weaknesses of
random generators~\cite{Dodis2013SAP}.  For instance, nonces generated
with weak pseudo-random generators can lead to attacks capable of
recovering secret keys.  Different security properties need to be
ensured when building strong pseudo-random generators (PRG),
such as resilience, forward security, backward security and recovering
security.  In particular, the latter was recently proposed as a
measure to recover the internal state of a PRG~\cite{Dodis2013SAP}.
\htext{
We propose a PRG that uses our source of strong entropy and implements
a refresh function to increase its resilience and recovery
capability.  
The pseudo-random generator
%, implemented by the Algorithm discussed in Appendix B of ~\cite{kreutz2017anchor}, 
has the following property:
}

%%%%%%%%%%%%%%
\textbf{Robust PRG -} Every value \textit{nprd} returned by the function
\textit{PRG\_next} is indistinguishable-from-random.

%\htext{
%A discussion about the correctness of the Algorithm can be found in Appendix B of ~\cite{kreutz2017anchor}.  
%}

A robust PRG needs three well-defined constructions, namely \textit{setup()}, \textit{refresh()} (or re-seed), and \textit{next()}, as described in Algorithm~\ref{alg:PRG}.
The internal state of our PRG is represented by three variables, the $seed$, the $counter$ and the next pseudo-random data $nprd$.
The setup process generates a new seed, by using our strong source of entropy, which is used to update the internal state.
In line 3, we initialize the $counter$ by calling the $long\_uint()$ function, which returns a long unsigned int value that will be used to re-seed and to generate the next pseudorandom value.
In line 4, we call $entropy\_update()$ to make sure that the external entropy gets updated before calling one more time the $entropy\_get()$ function.
The first $nprd$ is the outcome of an XOR operation between the newly generated seed and a second call to our source of entropy.
It is worth emphasizing that the set up of the initial state of the PRG does not require any intervention or interaction with the end user.
We provide strong and reliable entropy to set up the initial values of all three variables.
This ensures that our PRG is non-sensitive to the initial state.
For instance, in a tradicional PRG the user could provide an initial seed, or other setup values, that could compromise the quality of the generator's output.
The $counter$, which is concatenated with the $nprd$ (lines 9 and 13), gives the idea of an unbounded state space~\cite{stark2017dbet}.
This is possible because we are using cryptographically strong primitives such as a hash function H
and the MAC function HMAC.
Thus, in theory, we have unbounded state spaces, i.e., we can keep concatenating values to the input of these primitives.

\iftrue

\begin{figure}[ht]
  \centering
  \begin{minipage}{.5\linewidth}
  \begin{algorithm}[H]
	\caption{Pseudo-random generator}
	\label{alg:PRG}
	\begin{algorithmic}[1]
	\small
	\State \texttt{PRG\_setup()}
		\State \fourxspace \texttt{seed $\gets$ entropy\_get()}
		\State \fourxspace \texttt{counter $\gets$ long\_uint(entropy\_get())}
		\State \fourxspace \texttt{call entropy\_update()}
		\State \fourxspace \texttt{nprd $\gets$ seed $\oplus$ entropy\_get()}
	\vspace{2mm}
	\State \texttt{PRG\_refresh()}
		\State \fourxspace \texttt{seed $\gets$ entropy\_get()}
		\State \fourxspace \texttt{counter $\gets$ long\_uint(entropy\_get())}
		\State \fourxspace \texttt{nprd $\gets$ H(seed $\Vert$ nprd $\Vert$ counter)}
	\vspace{2mm}
	\State \texttt{PRG\_next()}
		\State \fourxspace \texttt{counter $\gets$ counter - 1}
		\State \fourxspace \texttt{if counter <= 0 call PRG\_refresh()}
		\State \fourxspace \texttt{\textbf{nprd}  $\gets$ HMAC(seed, nprd $\Vert$ counter)}
	\end{algorithmic}
  \end{algorithm}
  \end{minipage}
\end{figure}

\fi

The \textit{PRG\_refresh()} function updates the internal state, i.e., the $seed$, the $counter$ and the $nprd$.
It uses H to update the state of the $nprd$.
Finally, the \textit{PRG\_next()} function outputs a new, indistinguishable-from-random stream of bytes, applying HMAC  on the internal state.
In this function, the $counter$ is decremented by one. 
The idea is for it to start with a very large unsigned 8-bytes value, which is used until it reaches zero.
At this point, the  \textit{PRG\_refresh()} function will be called to update the internal state of the generator.
The newly generated $nprd$ is the outcome of an HMAC function with a dimension of 128 bits.

%One of the advantages of using a cryptographic HASH (line 8) is that the state space is unbounded, i.e., you can just keep concatenating new values.
%This unbounded state space is represented by the $u\_buffer$.
%Each time we call \textit{PRG\_refresh()} the unbounded buffer is incremented by one random byte from the entropy call $entropy\_get()$.
%The $nprd$ is updated (line 8) using a strong hash function (e.g., SHA512) which receives as input thee values, the newly generated $seed$, the current $nprd$ and $u\_buffer$.
% - one of the advantages of using a cryptographic HASH is that the state space is unbounded (you just keep adding, you concatenate one, two, etc. ; there is no limit; only the output is bounded;)
% What matters on the seed is the unpredictability! Entropy!
% u\_buffer provides an unbounded buffer 

\hdifftext{
The main motivation for having a PRG along with a strong source of entropy is speed.
Studies have shown that entropy generation can be rather slow, such as \SI{1.5}{\second} to \SI{2}{\minute} for generating 128 bits of entropy~\cite{Mahu2015SEG}.
Our source of entropy uses external entropy and random bytes from
special devices, whereas the PRG uses an HMAC function, in order to have a fast and reliable generation of pseudo-random values.
}

In spite of the fact that we could use any good PRG to generate cryptographic material 
(e.g. keys, nonce), it is worth emphasizing that we introduce a PRG that works 
in a seamless way with our strong source of entropy, improving its quality.
% which is a crucial requirement for having a good PRG.
In Section~\ref{sec:prg:gen}, we discuss the specifics of the implementation.
We also evaluate the robustness and level of confidence of our algorithms in Section~\ref{sec:PRG:entropy}.
%
%%%%%%%%%%%%%%%%%%%%%%% PROOFS %%%%%%%%%%%%%%%%%%
%
A discussion about the correctness of Algorithm~\ref{alg:PRG} can be found in Appendix B of ~\cite{kreutz2017anchor}.  %~\ref{appendix:PRG}.

%%%%%%%%%%%%%%%%%%%%%%%%%%%%%%
\subsection{Integrated device verification value (iDVV)}
\label{sec:iDVV}
%%%%%%%%%%%%%%%%%%%%%%%%%%%%%%

\hdifftext{
The design of our logically-centralized security architecture also includes the iDVV component~\cite{kreutz2016kiss}.  
The iDVV idea was inspired by the iCVVs (integrated card verification
values) used in credit cards to authenticate and authorize
transactions in a secure and inexpensive way. 
In~\cite{kreutz2016kiss} the concept was applied to SDN, proposing a flexible method of generating iDVVs that can be safely used to secure communication between any two devices. 
As a result, iDVVs can be used to partially address two gaps of non-functional properties, security-performance and complexity-robustness.
}

\ifarxiv

An iDVV is a unique value
generated by device A (e.g., forwarding device) which can be verified
by device B (e.g., controller). 
An iDVV generator has essentially two interfaces.  First,
\textit{idvv\_setup (seed, secret)}, which is used to set up the
generator.  It receives as input two secret, random and unique values,
the seed and the (higher-level protocol dependent) secret.  The source
of strong entropy and the robust PRG are, amongst other things,
used to bootstrap and keep the iDVV generators fresh.  Second, the
\textit{idvv\_next()} interface returns the next iDVV.  This interface
can be called as many times as needed.  

So, iDVVs are sequentially generated to authenticate and authorize
requests between two networking devices, and/or protect
communication. Starting with the same seed and secret, the iDVV
generator will generate, for example, at both ends of a
controller-device association, the exact same sequence of values.  In
other words, it is a deterministic generator of authentication or
authorization codes, or one-time keys, which are, however,
indistinguishable from random.  The main advantages of iDVVs are their
low cost, which makes them even usable on a per-message basis, and the
fact that they can be generated off-line, i.e., without having to
establish any previous agreement.

\else

\hdifftext{
iDVVs are sequentially generated to authenticate and authorize
requests between two networking devices, and/or protect
communication. Starting with the same seed and secret, the iDVV
generator will generate, for example, at both ends of a
controller-device association, the exact same sequence of values.  In
other words, it is a deterministic generator of authentication or
authorization codes, or one-time keys, which are, however,
indistinguishable from random.  The main advantages of iDVVs are their
low cost, which makes them usable even on a per-message basis, and the
fact that they can be generated off-line, i.e., without having to
establish any previous agreement.
}

\fi

\ifarxiv

\textbf{Correctness.} The randomness and performance of the iDVV
algorithm as deterministic generator of authentication or
authorization codes, or one-time keys which are however
indistinguishable from random, have been analyzed, and its properties
proved, in~\cite{kreutz2016kiss}.  The performance
study is complemented in Section~\ref{sec:iDVV:eval}.
Overall, these analyses show that iDVVs are robust,
achieve a high level of confidence and outperform traditional key
generation and derivation functions without compromising the security.

\else

\hdifftext{
The analysis provided in~\cite{kreutz2016kiss} and 
in this paper show that iDVVs achieve a high level of confidence 
and outperform traditional key generation functions without 
compromising the security.
}

\fi

%%%%%%%%%%%%%%%%%%%
\subsection{System roles and setup}
\label{sec:roles}
%%%%%%%%%%%%%%%%%%%

In our system we assume the existence of personnel with two different roles: the \emph{system administrator}, that controls
the operation of central services such as \ANCHOR, and the network
  administrator (a.k.a. manager), responsible for the operation of network devices. Every
time a new network device (a forwarding device or a controller) is
added to the network, it must first be registered, before being able
to operate. 

\hdifftext{
In the current practice, the device registration is a
manual process triggered by a network administrator through an
out-of-band channel.  
%This process would involve manual work from
%both the system and the network administrators. 
Given the potentially
large number of network devices in SDN, such a manual process is
unsatisfactory.
}
Thus, we propose a protocol, described below, to fulfil the desire of
a semi-automated device registration process, which is efficient,
secure, and requires the least involvement of \ANCHOR.  The \ANCHOR is
first set up by the system administrator.  
\htext{
Next, each network device is set up by \ANCHOR. 
Before that process, the network administrator has to share a secret key with the device.  
The set up of this key and the registration of devices is described in Section~\ref{sec:devreg}.
Then, the devices can be registered automatically.
}
% FIXME: Not clear. The net admin has to do something?

\hdifftext{
Now we present the deployment, communication and set up required for \ANCHOR (by the system administrator), network
  administrator, and devices.  Afterwards, we describe the device registration
  and association algorithms. 
}

\hdifftext{
\textbf{Deploying} \ANCHORBF.
Currently, \ANCHOR is designed to work in a single domain, with single ownership, such as a data center, enterprise, or university campus network.
\ANCHOR supports deployments with multiple controller instances~\cite{koponen2010}, for scalability and availability of network control, as is required in production systems~\cite{Jain2013BEG}.
It is worth emphasizing it is part of our plan to extend the \ANCHOR's features and services to multiple domains with multiple ownership.
}

\hdifftext{
\textbf{Connectivity infrastructure.}
\ANCHOR is designed to logically centralize non-functional properties of \emph{generic} SDN deployments.
As such, it is not restricted to OpenFlow. 
Other southbound APIs can be used, such as POF, ForCES, or P4.
The \ANCHOR connectivity infrastructure, used for communication between SDN devices (controllers and networking gear) and \ANCHOR, can use traditional in-band or out-of-band mechanisms (for instance, traditional routing protocols such as OSPF or IS-IS, as is common for control plane channels~\cite{koponen2010}).
%Messages from the device (e.g., controller, switches) to the root may pass through neighboring forwarding devices to reach \ANCHOR (and vice-versa). After successful registration, a forwarding device may connect to a controller recommended by \ANCHOR. Once the association between the devices has completed, the controller can select, adjust and even optimize the paths towards that forwarding device.
%In case of out-of-band Ethernet control networks, the process is similar but does not require to keep a topology three with \ANCHOR as its root.
}

\hdifftext{
For simplicity and without loss of generality, in what follows we
denote $E_{XY}$() an encryption using encryption key $Ke_{XY}$, and we
denote [],$\textit{HMAC}_{XY}$, respectively, a message field inside
[], followed by an HMAC over the whole material within [], using MAC
key $Kh_{XY}$, where $X,Y\in\{A,D_i,M,C,F\}$, for \emph{A}NCHOR, \emph{D}evice, network ad\emph{M}inistrator (or \emph{M}anager), \emph{C}ontroller, and \emph{F}orwarding device. 
%When $X=A$, we omit $X$ for simplicity. For example, we use $E_{M}$(msg) 
%to refer $E_{AM}$(msg), and they both denote the ciphertext of encrypting 
%$msg$ under key $Ke_{AM}$.
In what follows, \ANCHOR can generate strong keys using a suitable key
derivation function (KDF) based on the high entropy random material
described in the previous sections.
}
% FIXME: this paragraph is a bit confusing. Check Fernando's comments.

%FIXME: is this the best place to talk about the ANCHOR setup? Or move it to section 4.10?
\ANCHORBF \textbf{setup.}  The \ANCHOR needs two 
master recovery keys, namely the master
recovery encryption key $Ke_{rec}$ and master recovery MAC key
$Kh_{rec}$, fundamental for the post-compromise recovery steps
described ahead. However, these two master recovery keys, in
possession of the authority overseeing \ANCHOR (the system
administrator), must never appear in the \ANCHOR server (if they are
to recover from a possible full server compromise), being securely
stored and used only in an offline manner\,\footnote{Just to give a
  real feel, one possible implementation of this principle is: a
  pristine \ANCHOR server image is created; it boots offline in single
  user mode; it generates $Ke_{rec}$ and $Kh_{rec}$ through a strong
  KDF as discussed above; keys are written into a USB device, and then
  deleted; first online boot proceeds.}.
\htext{
Due to space constraints, we refer the reader to Appendix C of ~\cite{kreutz2017anchor} for more information (including a visual representation) regarding the three phases of \ANCHOR, namely setup, normal operation, and recovery.
}

As we will present later, the master recovery keys are only used in
two cases, namely (a) when a new network administrator is registered with
\ANCHOR (i.e. the network administrator setup process); and (b) when \ANCHOR
was compromised and is reinstated into a trustworthy state (i.e. the
post-compromise recovery process presented in
\S\ref{sec:post_compromise}). %For the occurrence of either case, 
When either case occurs, the 
\ANCHOR authority only needs to use the master recovery keys once, 
to recursively compute the recovery keys of all devices and network 
administrators.
The output of the calculation will be
imported into the \ANCHOR server through an out-of-band channel
(e.g. by using a USB).

\hdifftext{
\textbf{Network administrator setup.}  Each network administrator (or manager,
denoted $M$) with identity M\_ID is registered with \ANCHOR
manually. This is the \emph{only manual process} to initialize a new
$M$. Afterwards, all devices managed by this
M can be registered with \ANCHOR through our device
registration protocol.
}

During the network administrator registration phase, \ANCHOR locally generates
encryption key $Ke_{AM}$ and MAC key $Kh_{AM}$ to be shared with $M$,
and they are manually imported into $M$ through an out-of-band channel
(again, by using a USB, for example).

Further, $M$ recovery keys $Kre_{AM}$ = H($Ke_{rec}||\text{M\_ID}$) and
$Krh_{AM}$ = H($Kh_{rec}||\text{M\_ID}$) are also computed by \ANCHOR
offline. $M$ recovery keys live essentially offline, since $M$ needs
to perform only infrequent operations with these keys (e.g. upon
device registration).  Note that \ANCHOR does not store $Kre_{AM}$ or
$Krh_{AM}$ as well, but can recompute them offline when the
post-compromise recovery process is triggered, as we detail in 
Section~\ref{sec:post_compromise}.

\textbf{Device setup.}  A device with identity $D_i$ is either a
forwarding device (F) or a controller (C), but we do not differentiate
them during the set up and registration processes. 
The first operation to be made after a device is first brought to the
system is the setup, which, in the context of this paper, concerns the
establishment of credentials, for secure management access by the
network administrator.

Upon request from $M$, \ANCHOR locally generates a pair of keys for
each device $D_i$ being set up , $Ke_{MD_i}$ and $Kh_{MD_i}$, to be
respectively the encryption and MAC key to be shared between $M$ and
$D_i$, for management.  They are sent to $M$ under the protection of
$Ke_{AM}$ and $Kh_{AM}$. Then, they are manually imported by the
  network administrator into each $D_i$ through an out-of-band channel.
% (e.g. at the device's console, or by using a USB).

\subsection{Device registration}
\label{sec:devreg}

\highlighttext{
The device registration
protocol is presented in Algorithm \ref{alg:devReg}. We assume that $Ke_{MD_i}$ and
$Kh_{MD_i}$, described above, are in place.
}

\begin{figure}[ht]
  \centering
  \begin{minipage}{.75\linewidth}

\begin{algorithm}[H]
\footnotesize
\setstretch{1.3}
\nonumber
	\caption{Device registration}
	\label{alg:devReg}
	 \highlighttable
	\vspace{1mm}
% FIXME: XXXX
        \begin{tabular}{p{0.02\columnwidth}p{0.12\columnwidth}p{0.7\columnwidth}}
      
    &                                & \hspace{2em} \{Bootstrap for devices $D_1 - D_n$ \} \\\hline

1. & M $\rightarrow$ A & [$\text{Reg}, \text{M\_ID}, E_{M}$($\{(D_i,
                                x_m^i)\}_{i=1}^n$)], $\textit{HMAC}_{M}$ \\\hline

      2. & A & for each $D_i$, generate $Ke_{AD_i}, Kh_{AD_i}, x_a^i$\\
      3. & A $\rightarrow$ M & [$\text{Reg}, \text{M\_ID}, E_{M}$($\{$($D_i, x_m^i, x_a^i, Ke_{AD_i}, \allowbreak Kh_{AD_i}$)$\}_{i=1}^n$)], $\textit{HMAC}_{M}$ \\\hline
    &                                & \hspace{2em} \{For each device $D_i$\} \\\hline

      4. & M & $Kre_{{AD_i}} \gets$ H($Kre_{AM}||D_i$); $Krh_{{AD_i}} \gets$ H($Krh_{AM}||D_i$). \\\hline
      5. & M $\rightarrow D_i$ & [$\text{Reg}, E_{MD_i}$($x_a^i, Ke_{AD_i}, Kh_{AD_i}, \allowbreak Kre_{{AD_i}}, Krh_{{AD_i}}$)], $\textit{HMAC}_{MD_i}$ \\\hline
      6. & $D_i \rightarrow$ A & [$\text{M\_ID}, D_i, E_{D_i}$($x_a^i$)], $\textit{HMAC}_{D_i}$\\\hline
      7. & A $\rightarrow$ M & [$\text{M\_ID}, D_i, E_{M}$($x_a^i$)], $\textit{HMAC}_{M}$\\\hline
      8. & A & tag($D_i$) = registered;\\
      9. && for $t \in \{C,F\}$, if Type($D_i$)==t, then $tList = tList \cup \{Di\}$\\
      10. && $\forall i \in[1,n]$, if tag($D_i$) == registered is True\\
      11. && \hspace{1em} $Ke_{AM}$ = H($Ke_{AM}$); $Kh_{AM}$ = H($Kh_{AM}$).\\\hline
      12. & M $\rightarrow D_i$ & [$D_i, E_{MD_i}$($x_a^i$)], $\textit{HMAC}_{MD_i}$\\\hline
      13. & M & tag($D_i$) = registered;\\
      14. && destroys ($Ke_{AD_i}, Kh_{AD_i},Kr_{{AD_i}}$);\\
      15. && $Ke_{MD_i}$ = H($Ke_{MD_i}$); $Kh_{MD_i}$ = H($Kh_{MD_i}$);\\
      16. && $\forall i \in[1,n]$, if tag($D_i$) == registered is True\\
      17. && \hspace{1em} $Ke_{AM}$ = H($Ke_{AM}$); $Kh_{AM}$ = H($Kh_{AM}$).\\\hline
      18. & $D_i$  & $Ke_{MD_i}$ = H($Ke_{MD_i}$); $Kh_{MD_i}$ = H($Kh_{MD_i}$).\\\hline

     \end{tabular}
   \vspace{1mm}
\end{algorithm}

  \end{minipage}
\end{figure}

The first part concerns the bootstrap of the registration of a batch
of devices with \ANCHOR ($A$), by a network administrator $M$.
 Let $\{D_i\}_{i=1}^n$ be the set of $n$ device identities that the
 administrator wants to register.
$M$ requests (line 1) the registration to $A$, accompanying each $D_i$
 with a nonce $x_m^i$. $A$ computes its own nonce $x_a^i$, and keys
 $Ke_{AD_i}, Kh_{AD_i}$, for each $D_i$, and returns them encrypted to
 $M$ (lines 2,3).  The random nonces $x_m^i$ and $x_a^i$ are used to
 prevent replay attacks.

\htext{
The process then follows for each device $D_i$.  First, the device
recovery key is created (line 4), using $M$'s recovery keys $Kre_{AM}$ and $Krh_{AM}$.
}
Then $M$ sends $D_i$ the relevant cryptographic keys (line 5).
Device $D_i$ follows-up confirmation to $A$, which closes the loop
with $M$, using the original nonce from $A$ (lines 6,7).
$A$ then performs a set of operations (lines 8-11) to commit the
registration of $D_i$, namely by inserting it into the controller or
forwarding device list, respectively CList or FList, and updating
several keys.

Note that in Algorithm \ref{alg:devReg}, the update of several shared
keys (i.e., lines 11, 15, 17, 18) at the end of the
registration steps at $A$, $M$, and $D_i$, is used to provide PFS.
When a key is updated, the old one is destroyed.
Continuing, in line 12 $M$ closes the loop with $D_i$, using the
original nonce from $A$, finally confirming $D_i$'s registration.
Upon this step, both $M$ and $D_i$ perform the key update just
mentioned.

\htext{
Note that the generation process of the recovery keys $Krh_{AD_i}$ and $Krh_{AD_i}$ lies
with $M$ (line 4), though using its recovery keys shared with \ANCHOR,
$Kre_{AM}$ and $Krh_{AM}$. This reduces the number of uses of the master recovery
key. However, as we will see, albeit not knowing $Kre_{{AD_i}}$, $Krh_{{AD_i}}$, $Kre_{AM}$, and  $Krh_{AM}$,
\ANCHOR can easily compute them offline, if needed. Second, $Kre_{AM}$ and $Krh_{AM}$
possessed by the network administrator are only used when new devices need to
be registered. So, $Kre_{AM}$ and $Krh_{AM}$  can be usually stored offline. This
provides an extra layer of security.
}

%%%%%%%%%%%%%%%%%%%%%%%%%%%%%
\subsection{Device association}
\label{sec:devassoc}
%%%%%%%%%%%%%%%%%%%%%%%%%%%%%

The association service is required for authorizing control plane channels between any two devices, such as a forwarding device  and a controller.
A forwarding device has to request an association with a controller it
wishes to communicate with. This association is mediated by the
\ANCHOR.

The association process between two devices is performed by the
sequence of steps detailed in
Algorithm~\ref{alg:devAssoc}. Registered controllers and forwarding
devices are inserted in \textit{CList} and \textit{FList}, respectively.
\textit{Notation:} As explained above, the registration process set in
place shared secret keys between \ANCHOR (A) and any controller C or
forwarding device F.  

\begin{figure}[ht]
  \centering
  \begin{minipage}{.6\linewidth}
  \begin{algorithm}[H]
  \setstretch{1.3}
  \footnotesize
	\nonumber
	\caption{Device association} \hspace{1em} 
	\label{alg:devAssoc}
   % \highlighttable
	\vspace{1mm}
    \begin{tabular}{lll}
       &                                & \hspace{2em}  \{Of forwarding
          device $F$ with controller $C$\} \\\hline
    1. & F $\rightarrow$ A  & [$x_{g}$, F, GetCList],$\textit{HMAC}_{F}$   \\\hline
    2. & A $\rightarrow$ F  & [$x_{g}$, F, $E_{F}$(CList(F), $x_{g}$)],$\textit{HMAC}_{F}$   \\\hline
    3. & F $\rightarrow$ C & $x_{g}$, GetAiD, F, C, $E_{F}$(GetAiD, F, C, $x_{f}$, $x_{g}$)   \\\hline
    4. & C $\rightarrow$ A & [$x_{g}$, GetAiD, F, C, $E_{F}$(GetAiD, F, C, $x_{f}$, $x_{g}$),\\
        &                              &  $E_{C}$(GetAiD, F, C, $x_{c}$, $x_{g}$)],$\textit{HMAC}_{C}$ \\\hline
    5. & A $\rightarrow$ C & [$x_{g}$, $E_{F}$($x_{f}$, AiD), $E_{C}$($x_{c}$, AiD)],$\textit{HMAC}_{C}$ \\\hline
    6. & A & destroys ($AiD$)\\\hline 
    7. & C $\rightarrow$ F & $x_{g}$, $E_{F}$($x_{f}$, AiD), $E_{AiD}$(SEED, $x_{g}$)\\\hline 
    8. & F $\rightarrow$ C & $x_{g}$, $E_{AiD}$(SEED $\oplus$ $x_{g}$)\\\hline % shows  F knows  AiD and  SEED!
     9. & A, F & $Ke_{AF}$ = H($Ke_{AF}$); $Kh_{AF}$ = H($Kh_{AF}$)\\\hline 
  10.  & A, C & $Ke_{AC}$ = H($Ke_{AC}$); $Kh_{AC}$ = H($Kh_{AC}$)\\\hline 
     \end{tabular}
    \vspace{1mm}
\end{algorithm}

  \end{minipage}
\end{figure}

The device association implemented by Algorithm~\ref{alg:devAssoc}
has the following properties:

%%%%%%%%%%%%%%\begin{prop} 
\label{pprop1_algoAiD}
\textbf{Controller Authorization -} Any device F can only associate to a controller C authorized by the \ANCHOR.
%%\end{prop}

%%%%%%%%%%%%%%\begin{prop} 
\label{pprop2_algoAiD}
\textbf{Device Authorization -} Any device F can associate to some controller, only if F is authorized by the \ANCHOR.
%%\end{prop}

%%%%%%%%%%%%%%\begin{prop} 
\label{pprop3_algoAiD}
\textbf{Association ID Secrecy -} After termination of the algorithm, the association
ID ($AiD$) is only known to F and C.
%%\end{prop}

%%%%%%%%%%%%%%\begin{prop} 
\label{pprop4_algoAiD}
\textbf{Seed Secrecy -} After termination of the algorithm,  the seed
($SEED$) is only known to F and C.
%%\end{prop}

The algorithm coarse structure follows
the line of the Needham-Schroeder (NS) original authentication and key
distribution algorithm~\cite{Needham1978}, but contemplates anti-replay
measures such as including participant IDs, and a global initial nonce as
suggested in~\cite{Otway1987}. Unlike NS, it uses
encrypt-then-mac to further prevent impersonation. Furthermore, it is
specialized for device association, managing authorization lists, and
distributing a double secret in the end (association ID and seed).
Secure communication protocols running after association can, as
explained below in Section~\ref{sec:devcom}, use iDVVs on a key-per-message or
key-per-session basis, rolling from the initial seed and secret
association ID.

The association process starts with a forwarding device (F) sending an
association request to the \ANCHOR (A) (line 1 in
Algorithm~\ref{alg:devAssoc}).  This request contains a nonce $x_{g}$,
the identification of the device and the operation request $GetCList$
(get list of controllers).  The request also contains an HMAC to avoid
device impersonation attacks.  The \ANCHOR checks if F is in FList
(registered devices), and if so, it replies (line 2) with a list of
controllers (CList(F)) which F is authorized to associate with.  The
list of controllers (and the nonce $x_{g}$) is encrypted using a key
(set up during registration) shared between A and F.  This protects
the confidentiality of the list of controllers, and $x_{g}$ ensures
that the message is fresh, providing protection against replay
attacks.  A message authentication code also protects the integrity of
the \ANCHOR's reply, avoiding impersonation attacks.  Next, F sends an
association request to the chosen controller C (line 3).  The request
contains a message that is encrypted using a key shared between F and
A.  This message contains the get association id ($GetAiD$) request,
the identity of the principals involved (F,C), a nonce $x_{f}$, and
binds to the nonce $x_{g}$.  The controller forwards this message to A
(line 4), adding its own encrypted association request field, similar
to F's, but containing C's own nonce $x_{c}$ instead.
This
prevents the impersonation of the controller since only it would be
able to encrypt the freshly generated $x_{g}$. 
% FIXME: $x_{g}$? $x_{c}$?  %%PJV: _g, an impersonator could generate
% any _c, that's _g that makes the diff, IMHO
%%
In line 5, C trusts that
A's reply is fresh because it contains $x_{g}$.  The controller also
trusts that it is genuine (from A) because it contains $x_{c}$.  As
such, C endorses F as an authorized device and $AiD$ as the
association key for F.
\highlighttext{
Future compromise of A should not represent any threat to established 
communication between C and F. To achieve this goal, A immediately 
destroys the $AiD$ (line 6) and C and F further share a seed not known by 
A (line 7).
}

C forwards both the encrypted $AiD$ message and its seed to F (line
7).  The forwarding device trusts that this message is fresh and
correct because it contains $x_{g}$, and $x_{f}$ under encryption,
together with the $AiD$, only know to F and C, which it endorses then
as the association key.  F trusts that C is the correct correspondent,
otherwise A would not have advanced to step 5. That being true, future
interactions will use $AiD$.  F believes that the $SEED$ is genuine,
as random entropy for future interactions, because it is encapsulated
by $AiD$, known only to C and F.  The forwarding device also trusts
that the message is fresh because it contains $x_{g}$.  Finally (line
8), C trusts it is associated with F (as identified in step 3 and
confirmed by A in step 5), when F replies showing it knows both the
$AiD$ and the $SEED$, by encrypting the $SEED$ XOR'ed with the current
nonce $x_g$, with $AiD$.  Replay and impersonation attacks are avoided
because all encrypted interactions are dependent on nonces, so will
become void in the future. At the end of each device association
protocol, all keys shared between a device (F or C) and \ANCHOR will
be updated to the hash value of this key (lines 9, 10). Again, this is used to
provide perfect forward secrecy.  All nonces are random, i.e., not
predictable.

 \htext{
 A discussion of the correctness of Algorithm~\ref{alg:devAssoc} can be found in Appendix D of ~\cite{kreutz2017anchor}.  
 %Note that this algorithm is also formally verified using the \Tamarin prover, as discussed in Section~\ref{sec:security}.
 }

\ifarxiv

%%%%%%%%%%%%%%%%%%%%%%%%%%%%%
\subsection{Controller recommendation}
\label{sec:conrec}
%%%%%%%%%%%%%%%%%%%%%%%%%%%%%

Similarly to moving target defense strategies~\cite{Wang2014mtd}, devices (e.g., controllers) are hidden by default, i.e., only registered and authenticated devices can get information about other devices.
Even if a forwarding device finds out the IP of a controller, it will not be able to establish a connection with the controller unless it is registered and authorized by the \ANCHOR beforehand.

Controllers can be recommended to forwarding devices using different
parameters, such as latency, load, or trustworthiness.  When a
forwarding device requests an association with one or more
controllers, the \ANCHOR sends back a list of authorized controllers to
connect with.  The forwarding device will be restricted to associate
itself with any of the controllers on the list.  As a result,
forwarding devices will not be allowed to establish connections with
other (e.g., hostile or fake) controllers, and fake forwarding devices
will be, by default, forbidden to set up communication channels with
non-compromised controllers.

\fi

%%%%%%%%%%%%%%%%%%%%%%%%%%%%%
\subsection{Device-to-device communication}
\label{sec:devcom}
%%%%%%%%%%%%%%%%%%%%%%%%%%%%%

Communication between any two devices happens only after a successful
association. Consider the end of an association establishment, as per
Algorithm~\ref{alg:devAssoc}, e.g. between a controller C and a
forwarding device F: at this point, both sides, and only them, have the
secret and unique material $(SEED, AiD)$. %  (as proved in
% Appendix C of ~\cite{kreutz2017anchor}.
%~\ref{appendix:devAssoc}).  
Using them, they can bootstrap the
iDVV protocol (see Section~\ref{sec:iDVV} above), which from now on
can be used at will by any secure communication primitives.
As explained earlier, and detailed in~\cite{kreutz2016kiss}, iDVV generation is flexible and low cost, to allow the use: 
(a) on a per message basis; 
(b) for a sequence of messages; 
(c) for a specific interval of time; 
or (d) for one communication session. 

\hdifftext{
NaCl~\cite{Bernstein2012TSI}, as mentioned in previous
sections, is a simple, efficient, and secure alternative to
OpenSSL-like implementations, and is thus our choice for secure communication amongst devices.}
\ifarxiv
Indeed, researchers have shown that 
is resistant to side channel
attacks~\cite{Bacelar2013FV} and that its implementation is
robust~\cite{Bernstein2012TSI}.  Different from other cryptographic
libraries, NaCL's API and implementation is kept very simple,  
justifying its robustness.
%robust that it is (to date) impossible to use it in an incorrect way.
Through \ANCHOR, the SDN communication channels are securely encrypted
using symmetric key ciphers provided by NaCl, with the strong
cryptographic material required by the ciphers generated by our
logically centralized security mechanisms, allowing secret codes per packet, session, time interval,
or pre-defined ranges of packets.

\fi

%%%%%%%%%%%%%%%%%%%%%%%%%%%%%
\subsection{Post-compromise recovery}
\label{sec:post_compromise}
%%%%%%%%%%%%%%%%%%%%%%%%%%%%%

% FIXME: Section 4.10 – I think a statement is missing here regarding the recovery process i.e. the fact that ANCHOR re-instatement would only follow certain resolution of the compromise. Add some text to explain this.

\htext{
As previously explained, when \ANCHOR is reinstated after a compromise, it is crucial to have a way to automatically re-establish the secure communication channels between \ANCHOR and all participants.
}

Algorithm \ref{alg:recovery} presents our solution to re-establish the secure
communication channels when \ANCHOR is compromised.
Intuitively, since \ANCHOR's master recovery keys $Ke_{rec}$ and $Kh_{rec}$ are
stored securely offline, they are unknown to an attacker who
has stolen all secrets from the \ANCHOR server. As described before,
all $M$ and all $D_i$ recovery keys can be recursively computed from
the master keys, offline (line 1). 
\htext{
Afterwards, the system administrator imports those keys into the \ANCHOR server. To continue the recovery
process, \ANCHOR generates new random keys to be shared with all
$M$s, and all $D_i$ (line 2).
}

\htext{
Then, \ANCHOR sends to each $M$ (line 3) a recovery message to
re-share keys (contained in $M_k$) with the devices under the network administrator's control. 
The messages are encrypted with the corresponding recovery keys. 
The new shared keys will be used to protect future communications.  
Note that in line 3 we create an additional MAC value on the entire message
under the current MAC key $Kh_{AM}$.
Since the recovery keys are stored offline, without having this additional MAC value the
network administrator would have to perform the verification offline, manually. This
MAC value prevents possible DoS attacks where an attacker creates and
sends fake recovery messages to network managers, as this additional
MAC value can be verified online efficiently.
}

\htext{Each $M$ implements the recovery operation with each of
the devices it manages (line 4).}
The new keys replace the possibly compromised keys at $M$ and each
$D_i$ (lines 5-6, and 9).  Likewise, when the recovery process has
been completed, the recovery keys will be updated to their hash value
(lines 7-8, and 10-11).
As mentioned previously, this key update is
used to provide perfect forward secrecy (PFS).

%, and it cannot be created
%without having access to the current MAC key $Kh_{AM}$.

\begin{figure}[ht]
  \centering
  \begin{minipage}{.865\linewidth}
\begin{algorithm}[H]
\setstretch{1.3}
\footnotesize
\nonumber
	\caption{\ANCHOR recovery.}
	\label{alg:recovery}
    %\highlighttable
	\vspace{1mm}
    \begin{tabular}{lll}
       & & \hspace{2em}  \{For each manager $M$ and its associated devices $\{D_i\}_{i=1}^{n}\}$\}\\\hline
      1. & A & computes 
            $Kre_{AM}$ and $Krh_{AM}$% , $Kre_{AD_i}$, $Krh_{AD_i}$
               ;\\
      2. && generates 
$M_k=(Ke'_{AM}, Kh'_{AM}, \{Ke'_{AD_i}, Kh'_{AD_i}\}_{i=1}^{n})$.\\\hline
      3. & A $\rightarrow$ M & $[\text{Recovery}, \text{A, M\_ID}, [E_{Kre_{AM}}(M_k)], \textit{HMAC}_{Krh_{AM}}], \textit{HMAC}_{M}$.\\\hline
      & & \hspace{2em} \{For each device $D_i$\} \\\hline
      4. & M $\rightarrow$ $D_i$ & $[\text{Recovery}, \text{A, M\_ID, $D_i$}, [E_{Kre_{AD_i}}(Ke'_{AD_i}, Kh'_{AD_i})], \textit{HMAC}_{Krh_{AD_i}}], \textit{HMAC}_{MD_i}$.\\\hline
      5. & M  & destroys $Ke'_{AD_i}, Kh'_{AD_i}$;\\
      6. && $Ke_{AM} = Ke'_{AM}$; $Kh_{AM} = Kh'_{AM}$;\\
      7. && $Kre_{AM}$ = H($Kre_{AM}$); $Krh_{AM}$ = $H(Krh_{AM}$);\\
      8. && $Ke_{MD_i}$ = H($Ke_{MD_i}$); $Kh_{MD_i}$ = H($Kh_{MD_i}$).\\\hline
      9. & $D_i$ & $Ke_{AD_i} = Ke'_{AD_i}$; $Kh_{AD_i} = Kh'_{AD_i}$;\\
      10. && $Kre_{AD_i}$ = H($Kre_{AD_i}$); $Krh_{AD_i}$ = H($Krh_{AD_i}$);\\
      11. && $Ke_{MD_i}$ = H($Ke_{MD_i}$); $Kh_{MD_i}$ = H($Kh_{MD_i}$).\\
     \end{tabular}
    \vspace{1mm}
\end{algorithm}
\end{minipage}
\end{figure}

\hdifftext{
If the keys of the network administrator $M$ get compromised (e.g., if $M$ loses 
its keys), they can 
be recovered using the recovery keys provided by $A$.
Moreover, $M$ can also re-establish its shared secrets with \ANCHOR and 
its devices in a similar way as described in Algorithm~\ref{alg:recovery}.  
However, the steps are made only for a single $M$ instead of
all $M$, and with some differences, which we detail next.
First, $M$ gets the recovery keys (line
1) from \ANCHOR through an out-of-band channel: $Kre_{AM}$,
$Krh_{AM}$, and all $Kre_{AD_i}$, $Krh_{AD_i}$ from $i=1\ to\ n$. 
%These keys remain the same, but $M$ had lost them, having been rebuilt from scratch. 
Then, in lines 2-3, $M$ will get (generated by $A$) the $Ke'_{MD_i}, Kh'_{MD_i}$ keys
for managing all devices, instead of $Ke'_{AD_i}, Kh'_{AD_i}$, which
do not need to be changed.
Finally, in line 4 keys $Ke'_{MD_i}, Kh'_{MD_i}$ are sent to each
$D_i$, instead of $Ke'_{AD_i}, Kh'_{AD_i}$.}

% Security analysis
{\color{black}
%%%%%%%%%%%%%%%%%%%%%%%%%%%%%%
\section{Security Analysis}
\label{sec:security}
%%%%%%%%%%%%%%%%%%%%%%%%%%%%%%

We provide formal machine-checked verification of the core security
properties of \ANCHOR, using the \Tamarin~prover. In particular, we
formalise the core protocols of \ANCHOR, including device registration
protocol, device association protocol, and post-compromise recovery
protocol, through symbolic modeling. In addition, for each of the
protocols, we verify its correctness, message confidentiality, and
perfect forward secrecy (PFS). Moreover, we additionally verify the post-compromise security of \ANCHOR with the post-compromise recovery
protocol.

The full model contains 1712 lines of code. In total, we have proved
33 properties --- 23 of them are helper lemmas for the theorem prover
to understand \ANCHOR better; 4 lemmas are sanity proofs which check
the correctness of our protocols and their formalisation; and 6
main security properties that ensure the message confidentiality, perfect forward secrecy, and post-compromise security of \ANCHOR. We provide all input
files and complete formal model required to understand and reproduce our
security analysis at~\cite{tamarin-full-code-archive}.

% In particular,
% these include the complete model for the main \ANCHOR protocol,
% including device registration, device association, and post-compromise
% recovery.

\subsection{Security properties}

\ANCHOR achieves both classical security properties and novel security
properties. In a classical sense, the confidentiality of
communications between any two devices is guaranteed. In particular,
\ANCHOR also provides perfect forward secrecy, namely if a device is
compromised, then all communications of this device in the past are
still secure.

For the novel security guarantee, as mentioned before, rather than
assuming the trusted party cannot be compromised, such as CAs in X.509
PKI or the KDC in Kerberos, we also consider that \ANCHOR might be
compromised. In this case, we assume that there are means to detect
that the compromise has happened, and then the system can be recovered
through our post-compromise recovery protocol, which also guarantees
perfect forward security, when \ANCHOR is compromised and recovered.

\subsection{Formal analysis}
We analyse the main security properties of the protocol using the
\Tamarin~prover~\cite{Tamarin-cav}. The \Tamarin~prover is a symbolic
analysis tool that can prove properties of security protocols for an
unbounded number of instances and supports reasoning about protocols
with mutable global state, which makes it suitable for our
protocols. Protocols are specified using multiset rewriting rules, and
properties are expressed in a guarded fragment of first order logic
that allows quantification over timepoints.

\Tamarin is capable of automatic verification in many cases, and it
also supports interactive verification by manual traversal of the
proof tree. If the tool terminates without finding a proof, it returns
a counter-example.  Counter-examples are given as so-called dependency
graphs, which are partially ordered sets of rule instances that
represent a set of executions that violate the
property. Counter-examples can then be used to refine the model, and give
feedback to the implementer and designer.

\subsection{Modeling aspects}

As explained, we consider four protocol roles in \ANCHOR, namely A (\ANCHOR), M
(network Manager), F (Forwarding device), and C (Controller). To
simplify our model, we consider an additional role D (Device) to
represent any kind of network device, when it is irrelevant to distinguish its type (i.e., F or C).

We model the above protocol roles by a set of rewrite rules.
Our modeling of the roles follows the typical \Tamarin models, and directly corresponds to the algorithm descriptions in the previous
sections. 
Specifically, each rewrite
rule typically models receiving a message, taking an appropriate
action, and sending a response message.
\Tamarin provides built-in support for a Dolev-Yao style
network attacker, i.e., one who is in full control of the network. We
also specify rules that enable the attacker to compromise \ANCHOR
and/or any device in the network, and learn all of their session
keys.

\subsection{Proof goals}

We state several proof goals as specified in \Tamarin's syntax. Since
\Tamarin's property specification language is a fragment of
first-order logic, it contains logical connectives ({\tt |}, {\tt \&},
{\tt ==>}, {\tt not}, ...) and quantifiers ({\tt All}, {\tt Ex}).  In
\Tamarin, proof goals are marked as {\tt lemma}. The {\tt \#}-prefix
is used to denote timepoints, and ``{\tt E @ \#i}'' expresses that the
event $E$ occurs at timepoint $i$. Due to the space limitation, we
only present a set of examples selected from our full model, to
explain the core ideas. We refer the reader to the full model and detailed proof results available at
\cite{tamarin-full-code-archive}.

The first example goal is a check for executability that ensures that
our model allows for the successful transmission of a message. The
following example, which is a correctness lemma in the device
registration protocol, shows how it is encoded in our proof.

{\footnotesize
\begin{verbatim}
lemma protocol_correctness [use_induction]:
 exists-trace
  "Ex A D Did k keAD #i1.
     SendSec(A, D, Did,k, keAD) @ i1"
\end{verbatim}
}

The property holds if the \Tamarin model exhibits a behaviour in which
a device D of any type with unique identity Did can successfully
exchange with \ANCHOR A a message k encrypted by using a secret keAD
shared between D and A. This property mainly serves as a sanity check
on the model.  If it does not hold, it would mean our model does not
model the normal message flow, which could indicate a flaw in the
model. \Tamarin automatically proves this property and generates the
expected trace in the form of a graphical representation of the rule
instantiations and the message flow.
We additionally proved several other sanity-checking properties to
minimize the risk of modeling errors.

The second example goal is the core secrecy property with respect to a
classical attacker. When a controller C is associated with a
forwarding device F, then the following expresses that unless the
attacker compromises either C or F, he cannot learn any messages
exchanged between them. Note that {\tt K(m)} is a special event that
denotes that the attacker knows $m$ at this time.

{\footnotesize
\begin{verbatim}
lemma message_secrecy [use_induction]:
 "All C F Did1 Did2  k seed #i.
   /* If a message k is exchanged */
      ( SendSec(C, F, Did1, Did2, k, seed) @ #i &
   /* without the adversary compromising any device */
      not (Ex #j.
        Compromise_Device(C, F, Did1, Did2, seed) @ #j)
      ) ==>
   /* then the adversary cannot know k */
     not ( Ex #j. K(k) @ #j) "
\end{verbatim}
} \Tamarin also proves this property automatically. The above result
implies that if a forwarding device F with identity Did1 and a
controller C with identity Did2 has exchanged a message k encrypted
under a shared seed, and the attacker did not compromise any device
\emph{at any time}, then the attacker will not learn k.

Similarly, the following example expresses the PFS
for the communications between two devices.

{\footnotesize
\begin{verbatim}
lemma message_forward_secrecy [use_induction]:
 "All C F Did1 Did2  k seed #i.
      ( SendSec(C, F, Did1, Did2, k, seed) @ #i &
        not (Ex #j seed2.
         Compromise_Device(C, F, Did1, Did2, seed2) @ j &
        j<i)
      )
      ==>
      ( /* then the adversary cannot know k */
        not ( Ex #j. K(k) @ #j)
      )
 "
\end{verbatim}
}

\Tamarin proves this property automatically, and the result
additionally implies that the message is secure if the attacker did
not compromise any device \emph{before the current communication
  session.}

The final example property encodes the post-compromise security guarantees
provided by \ANCHOR. In this example, if \ANCHOR was compromised, and
then recovered through our protocol, then the confidentiality of
communications between \ANCHOR and forwarding device F is guaranteed.

{\footnotesize
\begin{verbatim}
lemma message_secrecy_after_recovery [use_induction]:
 "All A M F C Did k enckey #i1 #i2 #i3.
          (Comppromised_A(A) @#i1 &
           Recovery_Done(A,M,F,C)@ #i2 & i1<i2 &
           SendSec(A, F, Did, k, enckey) @ #i3 & i2<i3)
      ==>
     /* then the adversary cannot know k */
           not ( Ex #i4. K(k) @ #i4)
 "
\end{verbatim}
} The property states that if \ANCHOR was compromised at session $i1$,
and the recovery action has been completed afterwards at session $i2$,
then the confidentiality of message k exchanged in a later time
between A and forwarding device F is guaranteed.

\hdifftext{
The above properties are all proven automatically by the \Tamarin
prover on a PC\footnote{Intel(R) Core(TM) i7-6700 CPU @
  3.40GHz, 16GB memory.} within \SI{15}{\minute}.  
  }
 % Overall, the modeling
% effort was in the order of weeks, with several iterations to debug
% both the abstract model and the property specifications.  The
% verification process helped us to identify several flaws in our
% initial design, and to fix them according to the found traces.

}

% Security, yet simplicity
%%%%%%%%%%%%%%%%%%%%%%%%%%%%%%%%%%%%%%%%%%%%%%%%
\section{Implementation}
\label{sec:implementation}
%%%%%%%%%%%%%%%%%%%%%%%%%%%%%%%%%%%%%%%%%%%%%%%%

\diffminor{
A prototype of \ANCHOR has been implemented as envisioned in Figure~\ref{fig:architecture}.
To strengthen our confidence in the effectiveness of deployment of \ANCHOR in a production environment, we have implemented two versions of the system.
The first uses the POX controller and CBench\footnote{CBench is the default and most widely used tool for
  benchmarking control plane
  performance~\cite{khattak2014cbench,zhao2015cbench}.}
(OpenFlow switches emulator).
This version has approximately 2k lines of Python code and 700 lines of C code (integration with CBench). 
It uses Google's protobuf (\url{https://developers.google.com/protocol-buffers/}) for defining the protocols and efficiently serializing the data.
The second is a slightly simplified version using the Ryu controller and Open vSwitch.
In this section, we give an overview of the most relevant implementation details.  
The evaluation of the different components of the architecture is presented in Section~\ref{sec:evaluation}.
}
%performance~\cite{khattak2014cbench,zhao2015cbench}.}

%The \ANCHOR subsystem provides
%crucial root-of-trust mechanisms, source of strong entropy and PRGs,
%the iDVV protocol, and NaCl (a high-speed cryptographic
%library~\cite{Bernstein2012TSI}). We believe this is a promising
%alternative to PKI and OpenSSL.

%%%%%%%%%%%%%%%%%%%%%%%%%%%%%%
\subsection{Source of strong entropy}
\label{sec:strong:entropy}
%%%%%%%%%%%%%%%%%%%%%%%%%%%%%%

\hdifftext{
% entropy
We have 32 pools of events fed by four different sources, 
(1) incoming packet rate sent by controllers; 
(2) incoming packet rate of \ANCHOR; 
(3) network statistics of forwarding devices; and 
(4) random bytes from local systems.
Each of the sources feeds the pools in its own way. 
Sources (1) and (3) use a round-robin approach, whereas sources (2) and (4) randomly select the next pool to put the new event in.
In this way, we have a diversity of approaches for feeding the pools of noise, making the ``guessing task'' of an attacker harder.
Each pool needs to store only the digest of the SHA512 hashing function.
The current digest and the newly arrived events are used as input of the hashing function.
Lastly, once the pool has been used by the source of strong entropy, it is reset to a new initial state, which consists of the digest of a hash function using as input random bytes of a local entropy source such as \textit{/dev/urandom}.
}

%External sources of noise (e.g., forwarding device, controller) send heartbeats (a.k.a. events) to the \ANCHOR.
%Each heartbeat carries statistics of the current network traffic, idleness of links, and number of packets received by a controller within a specific time frame.

%\hdifftext{
%For setting up the external entropy \textit{entropy\_setup(data)} (see Algorithm~\ref{alg:entropyupdate}), the bytes read from the local source are combined (through an XOR operation) with the output of hashing function H($data$).
%We have chosen SHA512 as our strong hashing function $H$~\cite{dang2010recommendation}.
%After that, a second read of local random bytes is XORed with the external entropy to setup the internal entropy.
%}

\hdifftext{
To implement the \textit{entropy\_update()} function (see Algorithm~\ref{alg:entropyupdate}), we can use the pools of noise circularly (e.g., $P_0$ and $P_1$, $P_2$ and $P_3$, and so forth), in a combined circular and random way ($P_0$ and $P_7$, $P_1$ and $P_{31}$, and so forth), or in a completely random fashion.
The number of pools (32) and this diversity of approaches for using the pools make it very hard for an attacker to enumerate the possible values for the events used to update the generator's internal state~\cite{ferguson2011cryptography}.
}

\hdifftext{
Even if an attacker is controlling two or more external sources in a timely manner, it will be hard to guess the new state of the external entropy.
First, the attacker needs to enumerate the events of the pools being used on each update.
This, by itself, is something hard to achieve since the attacker does not know the update sequence of these pools, i.e., which external sources are being used, in which sequence, to update each pool.
In other words, he/she would have to know all sources of noise, and the sequence in which they are being used to update the pools.
It is also worth emphasizing that the external sources need to have a pre-defined maximum rate for sending the heartbeats, i.e., compromised sources cannot send data at a higher frequency to influence subsequent updates of the external entropy.
Second, the attacker would need to have additional knowledge regarding the internal entropy.
%, which is a result of two combined values, as explained in the following paragraph.
}
%, i.e., the two external sources being used at any given moment plus the output of the internal source.

%% entropy
%\textbf{Source of strong entropy}. 
%We implemented a source of strong entropy based on Algrithm~\ref{alg:entropyupdate}.
%It uses the Linux \textit{/dev/urandom} as a internal source of random bits.
%We have chosen SHA512 as our strong hashing function $H$~\cite{dang2010recommendation}.
%The pools of noise are used as sources of external entropy (i.e., $P_0$ and $P_1$).
%For each update of the external entropy, the events of two different pools compose the input of the SHA512 function.
%We use the pools in a circular fashion, i.e., we start with $P_0$ and $P_1$ and move on until we reach pools $P_{30}$ and $P_{31}$, starting from the beginning again.
%The resulting external entropy is an XOR between the output of the hash function and the current state of the internal entropy \textit{i\_entropy}.
%Finally, function \textit{entropy\_get()} outputs 64 indistinguishable-from-random bytes, which are the result of an XOR operation between the current state of the internal and external entropy.

%%%%%%%%%%%%%%%%%%%%%%%%%%%%%%
\subsection{Pseudorandom generator (PRG)}
\label{sec:prg:gen}
%%%%%%%%%%%%%%%%%%%%%%%%%%%%%%

\hdifftext{
Our PRG combines the implementation strengths
of different solutions such as the PRF of SPINS~\cite{Perrig2002SSP}
(which is based on an HMAC function), provably secure constructions
for building robust PRGs~\cite{Dodis2013SAP,ferguson2011cryptography},
and unbounded state spaces through cryptographic
primitives~\cite{stark2017dbet}.
}

As HASH function we have chosen SHA512.
As HMAC function, we have chosen the one time authentication function \textit{crypto\_onetimeauth()} from NaCl~\cite{Bernstein2012TSI}.
This function ensures security and performance while generating outputs of 16 bytes that are indistinguishable from random.

\hdifftext{
\textbf{PRG at the devices}.
As the controllers and forwarding devices do not have a source of strong entropy, we implemented a slightly modified version of the algorithm for these components to use this logically-centralized security service provided by the \ANCHOR.
Essentially, we replace the \textit{entropy\_get()} function by \textit{entropy\-\_remote()}.
Instead of using local data, this function makes an entropy request to the \ANCHOR to obtain a source of strong entropy.
This function is essential to provide recovering security by refreshing, improving the resilience of the PRG.
}

\subsection{Secure cryptographic key generators}
\label{sec:impl:iDVVs}
%%%%%%%%%%%%%%%%%%%%%%%%%%%%%

\hdifftext{
Based on the algorithm proposed in~\cite{kreutz2016kiss}, we have implemented an iDVV-based secure cryptographic keys generator that supports seven different cryptographic primitives.
Specifically, we use each of these primitives as input to the \textit{idvv\_next(primitive\_id)} function that is used to generate the next key.
In our implementation, we used the following primitives: \textit{MD5}, \textit{SHA1}, \textit{SHA512}, \textit{SHA256},  \textit{poly1305aes\_ authenticate}, \textit{crypto\_onetimeauth}, and  \textit{crypto\_hash}.
While the first four functions are provided by OpenSSL, the last three are provided by an independent implementation of Poly1305-AES and NaCl.
As MD5 and SHA1 are deprecated, we use them only for performance comparison purposes. 
}

\hdifftext{
To understand the rationale for our implementation, we give a bit of context to clarify the difference between our solution and traditional key derivation functions (KDFs).
Both solutions are used to generate secure cryptographic keys that can resist different types of attacks, such as exhaustive key search attacks~\cite{Yao2005dap}.
KDFs have common design characteristics, such as strong hash functions to compute digests for the raw key material.
A secure KDF can be defined as $H^{(c)}$($p||s||c$)~\cite{Yao2005dap}.
$H$ is a strong hash function such as SHA256 or SHA512. %~\cite{rfc6234}.
The exponent $c$ represents the number of iterations used to make the task of the attackers harder.
A common value for $c$ is $2^{16}$.
This exponent is particularly necessary if the entropy of the input $p$ (e.g., password, seed, key) is unknown.
In practice, the input of the KDF is likely to be of low-entropy~\cite{Yao2005dap}. %,Gennaro2003afp}.
While in some use cases a high exponent $c$ might be necessary to increase the cost of an attack trying to recover the key, it also significantly increases the cost of the key derivation function for high performance latency-sensitive applications.
}

\hdifftext{
Differently from a traditional key derivation scheme, our implementation using the iDVV generator in the context of \ANCHOR uses high-entropy values.
In other words, we do not need to recur to the exponent $c$ as a means to compensate a potentially low-entropy $p$.
By using by default two 32 bytes indistinguishable-from-random values in our generator, we make the task of an attacker very hard.
It is also worth mentioning that iDVVs are essentially used in an association basis, i.e., they have a relatively short lifetime.
}

%%%%%%%%%%%%%%%%%%%%%%%%%%%%%
\subsection{Implementation using Ryu and Open vSwitch}
\label{sec_impl_OpenvSwitch}
%%%%%%%%%%%%%%%%%%%%%%%%%%%%%

\diffminor{
We have implemented a second, simplified version of the system, focused on the essential registration and authorization functions of \ANCHOR.
We have used the Ryu controller~\cite{ryu2018component} for the control plane, and Open vSwitch (OVS) (\url{https://www.openvswitch.org/}) for the data plane. 
Ryu fully supports all versions of OpenFlow, including Nicira Extensions, and is officially integrated into OpenStack Networking (Neutron).
OVS is the main software switch used in virtualized data centers (e.g. VMWare NSX~\cite{vmware2018nsx}, OpenStack~\cite{openstack2018project}, OpenShift~\cite{redhat2018openshift}).
In addition, physical switches such as the Pica8 family~\cite{pica82018pica8} rely on PicOS~\cite{pica82018picos}, a user-space application running on top of an unmodified Linux kernel, providing OpenFlow support (version 1.0 to 1.4) through integration with standard Open vSwitch.
This second implementation further strengthens our case for the effectiveness of deployment of \ANCHOR in production SDN systems that include both software and hardware data planes.
}

\diffminor{
%Additionally, as shown by several previous research~\cite{lantz2010network,yan2015vt,de2014using,wang2014comparison,kaur2014mininet,fontes2015mininet,barrett2017dynamic,lantz2018mininet}, emulated OpenFlow-based networks, using Mininet~\cite{lantz2018mininet} and OVS, provide the requirements and the quality needed to realistically emulate and, afterward, deploy the exact same solutions in production environments.
}

% FIXME: improve this part
\diffminor{
We modified Open vSwitch (v2.10.0) and Ryu (v4.28) to support the registration and association functions provided by \ANCHOR. 
To evaluate our solution on a realistic scenario, using the Mininet emulator (\url{https://github.com/mininet/mininet})~\cite{lantz2010network,yan2015vt,de2014using,wang2014comparison,kaur2014mininet,fontes2015mininet,barrett2017dynamic} (further details in Section~\ref{sec_ovs_ryu_mininet}), we modified the behavior of the core hub module (\texttt{ryu/lib/hub.py}) of Ryu.
Similarly, we changed the behavior of the communication stack (\texttt{ovs/lib/stream}) of OVS.
Specifically, instead of just opening a new communication channel with the controller, our modified OVS registers itself with \ANCHOR (having obtained the network administrator's authorization), and sends an association request to the controller.
In Section \ref{sec_ovs_ryu_mininet} we present the results of \ANCHOR providing network protection against a rogue switch that is added to the network by an attacker, as an example use case.
}

%\diffminor{
%\textbf{TODO: write about the specifics of the implementation. stream-tcp.c/Open vSwitch, authentication and authorization before joining the Mininet topology, etc.}
%}

%Similarly to resilient PRGs~\cite{Dodis2013SAP}, the iDVV
%protocol also provides a synchronization mechanism that can be used to refresh the internal state of an %iDVV generator~\cite{kreutz2016kiss}.

%%%%%%%%%%%%%%%%%%%%%%%%%%%%%
%%%%%%%%%%%\subsection{ANCHOR}
%% PJV: this subsection did not make sense, deleted, text went to the
%% back, begin of sec 5
%%%%%%%%%%%%%%%%%%%%%%%%%%%%%

% Evaluation

%%%%%%%%%%%%%%%%%%%%%%%%%%%%%%%%%%%%%%%
\section{Evaluation}
\label{sec:evaluation}
%%%%%%%%%%%%%%%%%%%%%%%%%%%%%%%%%%%%%%%

In this section we evaluate the essential security mechanisms and services of our architecture.

For the performance measurements, we used machines with two quad-core Intel Xeon E5620 2.4GHz, with 2x4x256KB L2 / 2x12MB L3 cache, 32GB SDIMM at 1066MHz, with hyper-threading enabled.
These machines were interconnected by a Gigabit Ethernet switch and ran Ubuntu Server 14.04 LTS.

%%%%%%%%%%%%%%%%%%%%%%
\subsection{Source of entropy and PRGs}
\label{sec:PRG:entropy}
%%%%%%%%%%%%%%%%%%%%%%

We empirically evaluate both the source of strong entropy and PRGs through statistical methods and tools, following state of the art recommendations~\cite{Bassham2010SRS}.
To achieve our goal, we used NIST's test suite~\cite{nist2017sts}.
We generated one file containing 50MB of random bits per generator.
These files were used as input for the test suite tool STS~\cite{nist2017sts}.
In the end, our generators passed the absolute majority of tests and sub-tests: they failed only 2 sub-tests out of 188 (passed 146 out of 148 non-overlapping template matching), as summarized in Table~\ref{tabPRGs}.
This gives a very high level of confidence to our generators.

%{\renewcommand{\arraystretch}{1.3}
%\begin{table}[h!]
%\centering
%\footnotesize
%\begin{tabular}{ l | c } 
% \hline
% \textbf{Test} & \textbf{Result} \\ [0.5ex] 
% \hline\hline
%Frequency &  \checkmark \\\hline 
%Block Frequency & \checkmark \\\hline
%Cumulative Sums (forward) & \checkmark \\\hline
%Cumulative Sums (backward) & \checkmark \\\hline
%Runs & \checkmark \\\hline
%Longest Run of Ones & \checkmark \\\hline
%Binary Matrix Rank & \checkmark \\\hline
%Discrete Fourier Transform &  \checkmark \\\hline
%Non-overlapping Template Matching & 146/148 \\\hline
%Approximate Entropy & \checkmark \\\hline
%Random Excursions &  8/8\\\hline
%Random Excursions Variant & 18/18 \\\hline
%Serial (first) & \checkmark \\\hline
%Serial (second) & \checkmark \\\hline
%Linear Complexity & \checkmark \\\hline % [1ex] 
% \hline
%\end{tabular}
%\caption{STS: results of the single tests}
%\label{tabPRGs}
%\end{table}
%}

{\renewcommand{\arraystretch}{1.3}
\begin{table}[h!]
\centering
\footnotesize
\caption{STS: results of the single tests}
\label{tabPRGs}
    \begin{minipage}{.5\linewidth}
      %\caption{}
      \centering
\begin{tabular}{ l | c }
\hline
\textbf{Test} & \textbf{Result} \\ [0.5ex] 
\hline\hline
Frequency &  \checkmark \\\hline 
Block Frequency & \checkmark \\\hline
Cumulative Sums (forward) & \checkmark \\\hline
Cumulative Sums (backward) & \checkmark \\\hline
Runs & \checkmark \\\hline
Longest Run of Ones & \checkmark \\\hline
Binary Matrix Rank & \checkmark \\\hline
Discrete Fourier Transform &  \checkmark \\\hline
\hline
\end{tabular}
\end{minipage}%
    \begin{minipage}{.5\linewidth}
      \centering
        %\caption{}
\begin{tabular}{ l | c }
\hline
\textbf{Test} & \textbf{Result} \\ [0.5ex] 
\hline\hline
Non-overlapping Template Matching & 146/148 \\\hline
Approximate Entropy & \checkmark \\\hline
Random Excursions &  8/8\\\hline
Random Excursions Variant & 18/18 \\\hline
Serial (first) & \checkmark \\\hline
Serial (second) & \checkmark \\\hline
Linear Complexity & \checkmark \\\hline % [1ex] 
\textit{\# of sub-tests passed} & \textit{186/188} \\\hline % [1ex] 
\hline
\end{tabular}
\end{minipage} 
\end{table}
}

%%%%%%%%%%%%%%%%%%%%%%
\subsection{On the performance of key generation}
\label{sec:iDVV:eval}
%%%%%%%%%%%%%%%%%%%%%%

In this section, we analyse the performance of our key generator,
%%While the former increases the robustness and confidence in terms of security, the latter
which is essential to provide low latency and high throughput control plane communication at a low cost.
%We finish with a few words on implementation and application.

\hdifftext{
Figure~\ref{fig:results:iDVVgenerators} shows the latency of the seven cryptographic primitives we used with our generator.
We tested each primitive by generating keys of different sizes (16, 32, 64, and 128 bytes).
The best performance is achieved by the implementations based on SHA1 and MD5, as expected.
However, these two implementations have also the worst serial correlation coefficient, as shown in~\cite{kreutz2016kiss}.
The generators that use SHA512 or Poly-OTP have good performance, achieving a good security-performance tradeoff.
}

%\begin{figure}[ht]
%   \centering
%   \includegraphics[width=0.45\columnwidth]{idvv_funcs_bars_impl.pdf}
%   \caption{Latency of different iDVV generators}
%   \label{fig:results:iDVVgenerators}
%\end{figure}

%%%%%%%%%%%%%%%%%%%%%%
\subsection{Device-to-device communication performance}
%%%%%%%%%%%%%%%%%%%%%%

\textbf{Connection establishment.} While a TLS connection takes around \SI{19}{\milli\second} to be established, a device association using the \ANCHOR takes less than \SI{0.06}{\milli\second}.
\htext{
This means that \ANCHOR can easily support large-scale data centers (e.g., 1k switches and 100k hosts~\cite{greenberg2008towards,alfares2008scalable,benson2010network}) while being orders of magnitude more efficient than traditional solutions for this particular metric.
%This performance boost is a direct result of the fact that we use symmetric cryptography only, carefully selected high performance, yet secure, crypto primitives, and have less communication rounds when compared to TLS, for instance.
%Using \ANCHOR, the sum of the association time of all 800 switches is \SI{48}{\milli\second}.
%On the other hand, it takes \SI{15200}{\milli\second} using OpenSSL/TLS, i.e., approximately 300x more.
The scale of the improvement of our connection setup process when compared to the TLS handshake is due to three main factors.
First, our algorithm has half the number of steps.
Second, we use symmetric cryptography only.
Third, we use the fast ciphering provided by NaCl.
}

\begin{figure}
\begin{subfigure}{.5\textwidth}
  \centering
  \includegraphics[width=.85\linewidth]{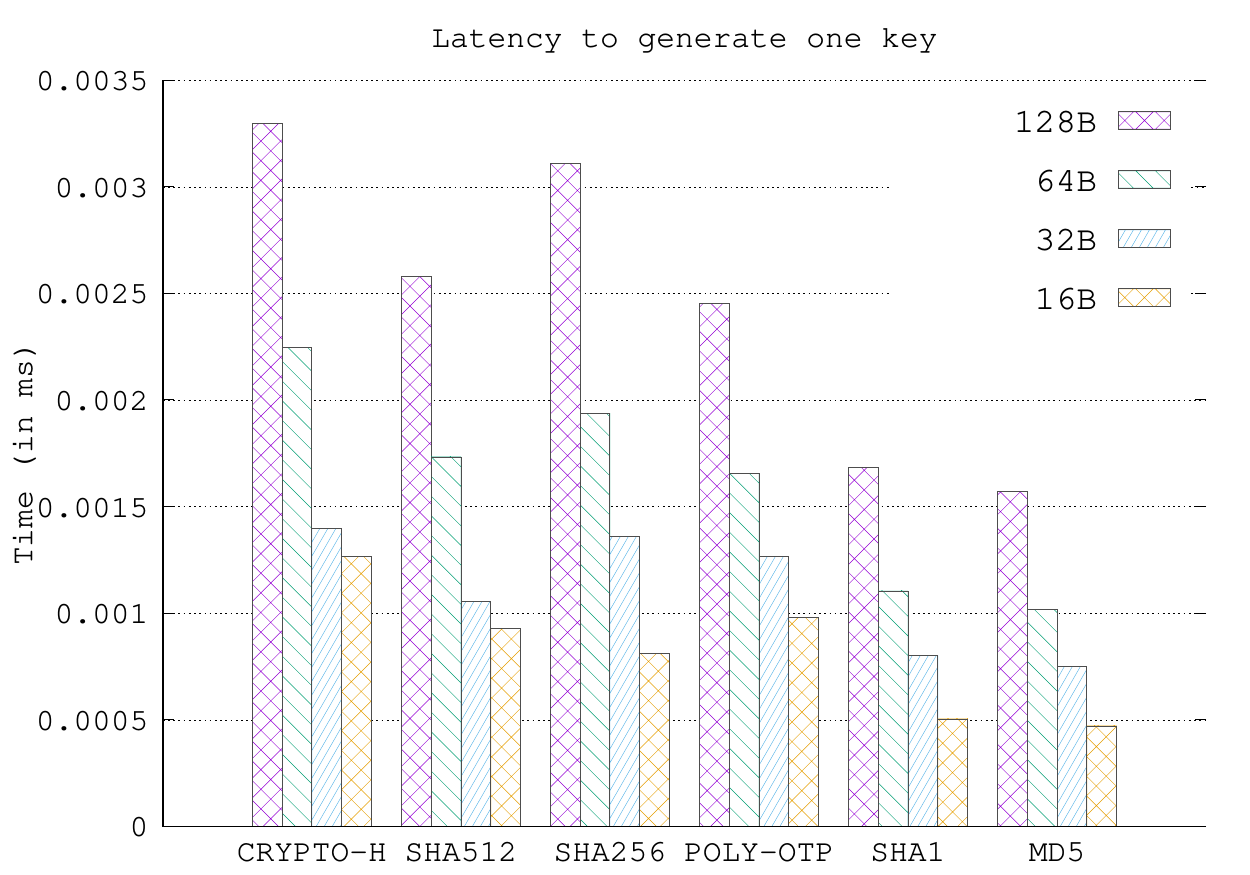}
  \caption{Latency of different key generators}
  \label{fig:results:iDVVgenerators}
\end{subfigure}%
\begin{subfigure}{.5\textwidth}
  \centering
  \includegraphics[width=.85\linewidth]{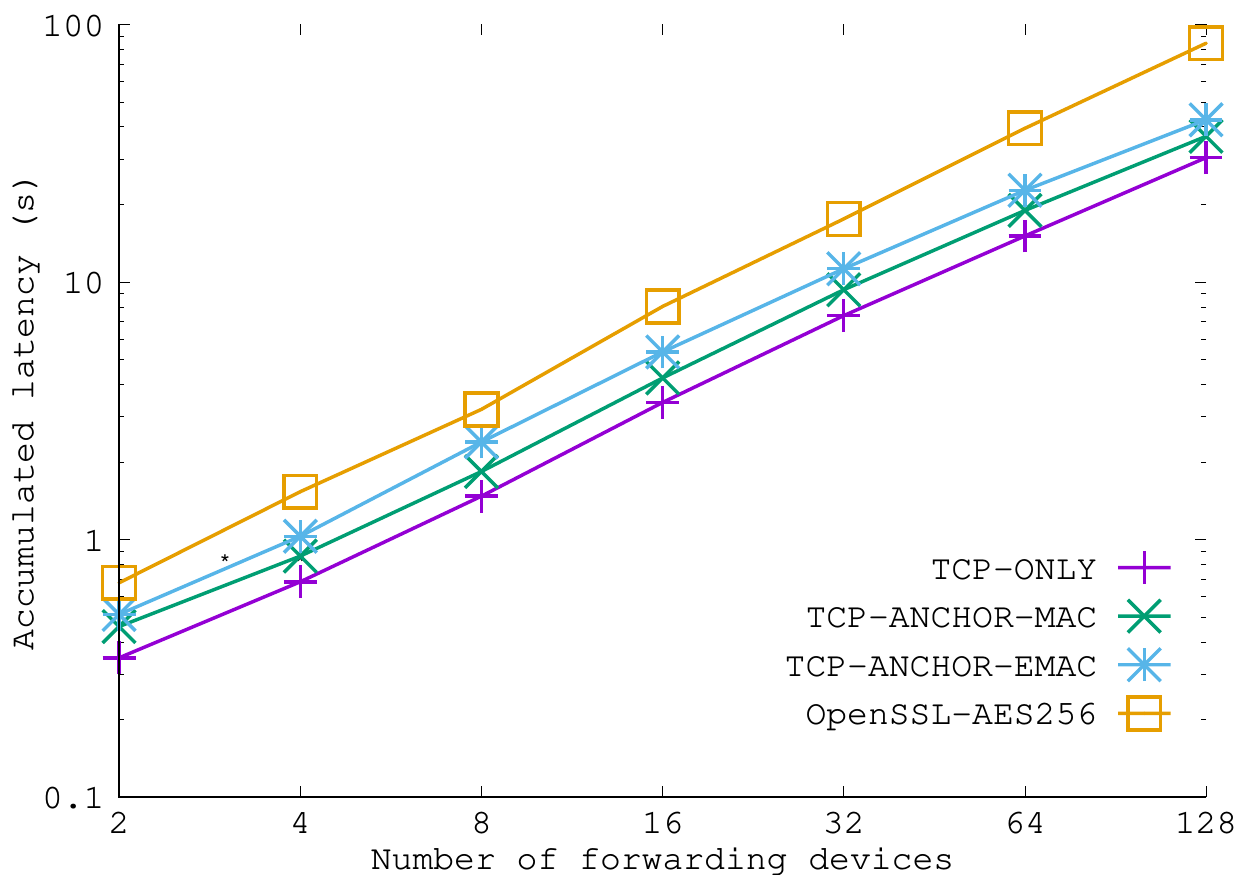}
  \caption{Control plane communication costs}
  \label{fig:results:comm:cost}
\end{subfigure}
\caption{Performance of key generation and control plane communications}
\label{fig:fig}
\end{figure}

\hdifftext{
\textbf{Communications overhead.} Figure~\ref{fig:results:comm:cost} shows the results of control plane communications using OpenSSL, TCP, and two versions of ANCHOR.
For communication of up to 128 forwarding devices, sending 10k control messages each, our solution requires (while offering stronger security guarantees - see below) only half of the resources and time of an OpenSSL-based implementation using AES256-SHA, the most widely available cipher suite. % --- adopted by most IT providers. % (e.g., Google, Facebook, Microsoft, and Amazon).
}

%As -of-band control channels are one example scenario where confidentiality of control plane communications might not be always required.
In Figure~\ref{fig:results:comm:cost}, we can also observe the overhead of confidentiality (TCP-ANCHOR-EMAC).
\htext{
In contrast to providing only authenticity and integrity (TCP-ANCHOR-MAC), confidentiality incurs an overhead of around 15\%.
}

It is worth emphasizing that we achieved these results by ensuring also much stronger security, as we generated one secret key \emph{per packet}.
On the other hand, the OpenSSL-based implementation used a single key (for the symmetric ciphering) for the entire communication session.

%\begin{figure}[ht]
%   \centering
%   \includegraphics[width=0.45\columnwidth]{comm_tcp_idvv_openssl_v2.pdf}
%   \caption{Control plane communication costs}
%   \label{fig:results:comm:cost}
%\end{figure}

%\begin{figure}[ht]
%   \centering
%   \includegraphics[width=0.45\columnwidth]{idvv_funcs_bars_impl.pdf}
%   \caption{Latency of different iDVV generators}
%   \label{fig:results:iDVVgenerators}
%\end{figure}

%%%%%%%%%%%%%%%%%%%%%%
\subsection{Attack prevention}
\label{sec_ovs_ryu_mininet}
%%%%%%%%%%%%%%%%%%%%%%

\diffminor{
A type of attack that is recurrently presented as an important security threat in the context of SDN is the introduction, by an attacker, of rogue switches in the network (see~\cite{kreutz2013HotSDN,antikainen2014spook,chi2015detect,kamisinski2015flowmon,chiu2017rapid,li2016survey}).
A set of switches under control of an attacker can be used for a DDoS attack, for instance, negatively affecting SDN control.
We use this type of attack as an example use case that shows the logical centralization of security services in \ANCHOR to enable attack prevention. 
The defence against this type of attack consists of a switch having to register itself to \ANCHOR before being able to associate with the controller.
If either the registration or association process fails, the switch connection with the controller is automatically dropped.
}

\diffminor{
To demonstrate this functionality, we set up an experiment using the second version of our system (the one with OVS and Ryu as the data and control planes, respectively).
We emulated a small network with Mininet, comprised of five switches (s0 to s4) and five hosts (h0 to h4), following a tree topology with s0 at the root.
Each network host is connected to one switch (e.g., host h4 is connected to switch s4).
To emulate the attack, we assumed s2 to be a rogue device introduced to the network by an attacker.
As the network manager has not registered s2 into the system, this switch should not be able to associate itself with the controller.
As a result, host h2 should not be reachable by any other host, and vice-versa.
%In addition to the modifications made in Ryu (see Section \ref{sec_ovs_ryu_mininet}), we used an app that emulates the behavior of a standard Ethernet switch.
}

\diffminor{
The outcome of the experiment was as follows.
Once the mininet network was up and running, we have run the \texttt{pingall} command to verify the reachability of all hosts.
We observed an overall packet loss of 40\% -- the result of 1 out of 5 unreachable hosts (h2, in this case). 
Each host executes a reachability test for all other four hosts.
However, the reachability test for host h2 fails. 
In case of h2, all reachability tests fail.
In a closer inspection, we verified that while the simulation was running, switch s2 periodically tried to associate itself with Ryu, without success, as expected.
%However, as s2 has no authorization from \ANCHOR, the controller just keeps dropping the association requests.
}

%*** Creating network
%*** Adding controller
%Connecting to remote controller at 127.0.0.1:6653
%*** Adding hosts:
%h0 h1 h2 h3 h4 
%*** Adding switches:
%s0 s1 s2 s3 s4 
%*** Adding links:
%(h0, s0) (h1, s1) (s1, s0) (s2, h2) (s2, s0) (s3, h3) (s3, s0) (s4, h4) (s4, s0) 
%*** Configuring hosts
%h0 h1 h2 h3 h4 
%*** Starting controller
%c0 
%*** Starting 5 switches
%s0 s1 s2 s3 s4 ...
%*** Starting CLI:
%mininet> pingall
%*** Ping: testing ping reachability
%h0 -> h1 X h3 h4 
%h1 -> h0 X h3 h4 
%h2 -> X X X X 
%h3 -> h0 h1 X h4 
%h4 -> h0 h1 X h3 
%*** Results: 40% dropped (12/20 received)
%mininet> 

%%%%%%%%%%%%%%%%%%%%%%
\subsection{Traditional solutions $versus$ ANCHOR}
\label{sec:evaluation:versus}
%%%%%%%%%%%%%%%%%%%%%%

In Table~\ref{tab:tradvour} we provide a summarised comparison between a traditional solution and our \ANCHOR.
As traditional solutions we considered the EJBCA (\url{http://www.ejbca.org/}) and OpenSSL, two popular implementations of PKI and TLS, respectively.
As we have shown before, our bootstrap process (device registration and association) is much faster and our connection latency is also significantly lower.
\hdifftext{
In addition, our solution has nearly one order of magnitude less LOC and uses four times fewer external libraries. This makes a difference from a resilience perspective. 
}
For instance, to formally prove more than 717k LOC (EJBCA + OpenSSL) is by itself a tremendous challenge. 
\htext{
Moreover, it gets considerably worse if we take into account eighty external libraries and eleven programming languages.
}

{\renewcommand{\arraystretch}{1.2}
\begin{table*}[!htp]
\caption{Traditional solutions $versus$ \ANCHOR}
\label{tab:tradvour}
%\highlighttable
\begin{center}
\footnotesize
%\rowcolors{1}{lightgray}{white}
\begin{tabularx}{\linewidth}{>{\tablefontsize}p{2.2cm}>{\tablefontsize}p{1.4cm}>{\tablefontsize}X>{\tablefontsize}X}
\hline
\textbf{Functionality} & & \textbf{Traditional solutions} & \ANCHORBF \\\hline

Typical Software 
&  & EJBCA (PKI) + OpenSSL (TLS) & \ANCHOR + iDVV + NaCl \\\hline

Device Identification
&  & based on certificates; costs = issue a certificate & based on unique IDs controlled by the \ANCHOR; costs = register the device (assign a unique ID) \\\hline

Device Registration
&  & based on certificates; costs = certificate installation + security control policy/service & registration protocol; costs = register the device + iDVV bootstrap \\\hline

Device Association \& KeyGen
& & 12 step mutual handshake + DH for session keys (incl. certificate validation - any two device can establish an association) & 6 step trust establishment with \ANCHOR + iDVVs per message, session, interval of time, ... (\ANCHOR has to authorize association) \\\hline

\multirow{5}{*}{Security Properties} 
& Authenticity      & \repeatvalue{8}{\whitehspace} \cmark  & \repeatvalue{8}{\whitehspace}\cmark   \\\cline{2-4}
& Integrity            &\repeatvalue{8}{\whitehspace} \cmark  & \repeatvalue{8}{\whitehspace}\cmark   \\\cline{2-4}
& Confidentiality  & \repeatvalue{8}{\whitehspace} \cmark  & \repeatvalue{8}{\whitehspace}\cmark   \\\cline{2-4}
& PFS  & \repeatvalue{8}{\whitehspace} \cmark(*)  & \repeatvalue{8}{\whitehspace}\cmark   \\\cline{2-4}
& PCS  & \repeatvalue{8}{\whitehspace} \xmark  & \repeatvalue{8}{\whitehspace}\cmark   \\\cline{2-4}
& PQS  & \repeatvalue{8}{\whitehspace}  \xmark  & \repeatvalue{8}{\whitehspace}\cmark   \\\cline{2-4}
\hline

Communications 
&  & symmetric cryptography (cipher: AES256-SHA) & symmetric cryptography (cipher: Salsa20) \\\hline

TLS stack 
&  & highly configurable and complex (717k LOC) & easy to use, simple (85k LOC), and efficient \\\hline

\end{tabularx}
\end{center}
\end{table*}
}

Our proposed architecture offers a functionally equivalent level of security with respect to properties such as authenticity, integrity and confidentiality, when compared to traditional alternatives.
Additionally, \ANCHOR offers a higher level of security by providing post-compromise security (PCS) and post-quantum security (PQS).
While the former is ensured through post-compromise recovery (see Section~\ref{sec:post_compromise}), the latter is a consequence of using only symmetric cryptography.
Further, the lightweight nature of our mechanisms, such as the iDVV, make them amenable to be used on a per message basis to secure communication, increasing cryptographic robustness.
\htext{
Moreover, by having fewer LOC, we significantly reduce the threat surface. 
}

Finally, it is worth emphasizing that the perfect forward secrecy (*) of traditional solutions, such 
as those provided by the different  implementations of TLS, is not easy or simple to enforce.
First, in spite of TLS providing ciphers that offer PFS, in practice, different cipher suites do 
not feature it~\cite{rfc7525}. This means that not all implementations and deployments of TLS 
offer PFS, or provide it with very low encryption grade~\cite{huang2014aes,digicert2017epf,namecheap2015csc}.
\hdifftext{
To give an example, widely deployed web  servers, such as Apache and Nginx~\cite{digicert2017epf} and most DHE- and 
ECDHE-enabled servers suffer from weak PFS configurations~\cite{huang2014aes,adrian2015ifs,springall2016msh}.}

% Related work
%%%%%%%%%%%%%%%%%%%%%%%%%%%%%%%%%%%%%%%%%%%%%%%%%%%%%%
\section{Related work on SDN security}
\label{sec:rwork}
%%%%%%%%%%%%%%%%%%%%%%%%%%%%%%%%%%%%%%%%%%%%%%%%%%%%%%

% ~\cite{tatlicioglu2018method}

% ADD: KISS, ONOS, RADIUS, ...

\diffminor{
Related work on SDN security (see~\cite{hayward2016asurvey,kreutz2014sdnsurvey,dacier2017sco,yoon2017flow} for broad surveys) focuses on securing \emph{specific} components of the architecture.   
As many attacks exploit vulnerabilities of the control plane, the security of the controller and the applications running on top of it have deserved good attention.
In a nutshell, solutions such as SANE~\cite{casado2006sane}, Ethane~\cite{Casado2007}, FortNOX~\cite{porras2012sek}, SE-Floodlight~\cite{porras2015securing}, FRESCO~\cite{shin2013fresco}, FLOWGUARD~\cite{hu2014flowguard}, are specialized applications that run on top of (or as) controllers. 
The main goal of such solutions is to provide some security enforcement in the data plane through security services that run on the control plane.
%conflict resolution of rules generated by different security apps, and the enforcement of isolation and priority between different types of apps running on top of controllers.
For instance, FLOWGUARD~\cite{hu2014flowguard} allows users to build robust firewalls, by means of a comprehensive framework, to protect OpenFlow-based networks. 
Those firewalls generate OpenFlow flow rules to be installed in forwarding devices, protecting devices and network services against different security threats.
As another example, the controller Rosemary~\cite{shin2014rosemary} implements a network application containment and resilience strategy that addresses the problem of malicious applications leading to loss of network control.  
Similarly, FortNOX~\cite{porras2012sek}, a software extension for the NOX controller, is robust to adversarial applications by providing role-based authorization and security constraint enforcement.
}

\diffminor{
These works address different security issues, and they all focus on the functional aspects: i.e., on installing the appropriate OpenFlow rules in the data plane to achieve their goals. 
In contrast, we propose the logical centralization of \emph{non-functional properties}, with focus on infrastructure security in this paper.
%For instance, the security of device registration and association, assessment of trust among devices, and high-quality cryptographic material. 
As such, \ANCHOR should be seen as: (i)  complementary to these solutions; and (ii) in fact, providing overarching properties which can, amongst other things, assist in the robust implementation of some of the proposed services. For instance, Rosemary requires a PKI infrastructure for application authorisation that could be replaced by \ANCHOR, inheriting its advantages.
}

\htext{Another line of work in SDN security is devoted to DoS/DDoS attack detection and prevention.
As an example, the use of lightweight information hiding based authentication (by means of secrecy through obscurity) has been proposed as one way of protecting SDN controllers from this type of attack~\cite{abdullaziz2016light}.
The idea is to use a specific field in the IP protocol to hide the switch authentication ID. 
In order for the scheme to be workable, it is assumed that a look-up table and unique IDs are shared among devices through existing key distribution protocols.
Again, this point solution could take advantage of \ANCHOR for this purpose.}
%While such lightweight technique can indeed be used to mitigate DoS attacks, it does not address the security issues of control plane communications -- such as authenticity, integrity, confidentiality, and data freshness -- we address here.

\htext{Interestingly, not much attention has been paid to the security of control plane associations and communication between devices, one of the aspects we address in this paper.
%It is typically assumed that TLS can be used to secure the communication between forwarding devices and controllers.
While TLS is the solution recommended by ONF, recent research discusses the strengths and weaknesses of this protocol as a means to provide authenticated and encrypted control channels~\cite{samociuk2015secure}, which is aligned with many of the arguments we make here. 
As we explained, while the use of TLS gives important security properties, it has an impact on control plane performance.
Additionally, the complexity of the infrastructure software has been recurrently pointed out as one of the main causes for a high number of reported vulnerabilities, that in many cases have led to security attacks~\cite{Zhou2012dam,Markowsky2013wt2,McGraw2004sse,hoepman2007ist}.
As we argue in this paper, by logically-centralizing crucial security mechanisms, \ANCHOR removes complexity from both controllers and switches, enhancing the robustness of the infrastructure, and still achieving a gain in performance.}

% Lightweight Authentication Mechanism for Software Defined Network using Information Hiding~\cite{abdullaziz2016light}

%\htext{
%Similarly, other solutions have been proposed to address specific issues of OpenFlow-enabled SDNs, such as the duplication of data path IDs~\cite{kang2015mynah,park2017sdn} and user authentication in control plane applications (e.g., back-end authentication and authorization services such as OpenLDAP and FreeRADIUS for OpenDaylight)~\cite{odl2018security,frinxio2018frinx}.
%However, these solutions do not go beyond specific security issues.
%For instance, LDAP and RADIUS back-ends are used just for user authentication and authorization in OpenDaylight, which is what these protocols were designed for.
%OpenDaylight (as well as ONOS and other OpenFlow controllers) uses TLS for providing integrity, confidentiality, and authenticity of control plane communications between the %controller and switches.
%However, it does not provide any security properties in communications between controller instances, i.e., controller-to-controller communications.
%As we discussed before (see Section~\ref{sec:background}), TLS, by itself, might not be enough to provide high-performance control plane communications. Additionally, TLS implementations might not ensure PFS (see Section~\ref{sec:evaluation:versus}).
%}

\htext{
Finally, to protect control plane communications between controllers and forwarding devices our solution makes use of two existing mechanisms: iDVV~\cite{kreutz2017kiss,kreutz2016kiss}, as a secure and low-cost method for generation of authentication codes, and NaCl~\cite{Bernstein2012TSI}, as a robust alternative to OpenSSL.  
We apply these solutions to SDN, but given their standalone nature they can be applied to different scenarios.
}

%However, what we do at \ANCHOR is more ambitious (a fully fledged
%architecture) and has architectural reach where iDVV is just a minor
%piece.

\diffminor{
To our knowledge, an architectural approach as the one we propose here  (which ultimately led to following the SDN philosophy of ``logical centralization'') was lacking.
Importantly, this approach allowed us to gain a global perspective of the relevant gaps in SDN and the limitations of existing solutions to the problem. 
This first step gave insight into one of the most relevant problems of SDN (as noted by the ONF or MEF security groups~\cite{onf2017onf,mef2017mef}): the security of the associations and communications between devices -- which jointly with the architecture itself, is one of the contributions of our paper.
}

%%%%%%%%%%%%%%%%%%%%%%%%%%%%%%%%%%%%%%%%%%%%%%%%%%%%%%
\section{Discussion}
\label{sec:discussion}
%%%%%%%%%%%%%%%%%%%%%%%%%%%%%%%%%%%%%%%%%%%%%%%%%%%%%%

\hdifftext{
We briefly discuss how we filled the gaps identified in
Section~\ref{sec:background}.
%, with our specialization of the logically centralized \ANCHOR architecture for `security'. 
Incidentally, we also show, in Appendix E of ~\cite{kreutz2017anchor}, %~\ref{appendix:secReq},
to which extent these solutions cover eleven of ONF's security requirements.
We conclude the section with a critique of our choices and results.
}

%%%%%%%%%%%%%%%%%%%%%%%%%%%
\subsection{Meeting the challenges}
%%%%%%%%%%%%%%%%%%%%%%%%%%%

\hdifftext{
\textit{Security $vs$ performance?}  Control channels need to provide
high performance (high throughput and low latency) while keeping the
communication secure.  However, as it has been shown, security
primitives have a non-negligible impact on performance.  To mitigate
this problem, we selected appropriate cryptographic
primitives (SHA512), libraries (NaCl), and key generation mechanisms (iDVV) to
ensure the security of control plane communications maintaining high performance.
%Additionally, the proposed iDVV allows systematic refreshing of crypto material with high performance, while further improving cryptographic robustness.
By logically centralizing the fundamental aspects of these mechanisms in the \ANCHOR, the performance overhead introduced in forwarding devices and controllers is limited, as they require only minimal functionality to `hook' to the \ANCHOR instructions.    
}

\textit{Complexity $vs$ robustness?}
\htext{
Traditional implementations of SSL/TLS, such as OpenSSL, have a large, complex code base, that recurrently leads to
vulnerabilities being discovered.
}
Similar problems are observed in PKI subsystems.  
It is well know that an effective means to achieve robustness is by reducing complexity.
\htext{
Hence our choice for the NaCl and iDVV mechanisms to help fill the gap, since they
are respectively lightweight (small code base), efficient, yet secure alternatives to
OpenSSL-like implementations.
}
\hdifftext{
As such, they are a robust solution to provide authentication and authorisation material for the secure
communications protocols we propose.
They are also amenable to verification mechanisms aimed to assure correctness, which are much harder to employ in very large code bases.
Again, the centralization of the non-functional mechanisms introduced in our solution is the key to reduce complexity of networking devices, improving their robustness.  
}

\hdifftext{ \textit{Global security policies?}  We have argued that controllers and
network devices often employ suboptimal network authentication and secure
communication mechanisms, despite recommendations from ONF and other such organizations for the opposite.
This problem is very similar in nature to the state of affairs in networking before SDN.
In traditional networks, enforcing relatively ``simple'' policies such as access control rules~\cite{Casado2007} or traffic engineering mechanisms~\cite{Jain2013BEG} was either very hard or even impossible in practice. 
Given the current undesirable 
state of affairs, we believe the same to be true to non-functional properties, with security as a prominent example.
Our logically centralized \ANCHOR architecture addresses this gap by providing a means for making centralized policy rules 
%(e.g., about registration, authentication and association of network devices) 
and the necessary mechanisms to enforce them, permeating the SDN architecture in a global and coherent way.
}

\textit{Resilient roots-of-trust?}  We debated that there is a
(probably reduced) number of functions which should not be left to ad-hoc
implementations, due to their criticality on system correctness.  The
list is not closed, but we hope to have shown that strong sources of
entropy and resilient indistinguishable-from-random number generators are clear examples
of difficult-to-get-right mechanisms that benefit from such logically centralized approach.
\ANCHOR addresses this issue, by providing the motivation to design and verify careful and
resilient once-and-for-all implementations of such root-of-trust
mechanisms, which can then be reinstantiated in different SDN
deployments.

%%%%%%%%%%%%%%%%%%%%%%%%%%%
\subsection{Devil's advocate analysis}
%%%%%%%%%%%%%%%%%%%%%%%%%%%

\iftrue

\textit{Doesn't the logical centralization of non-functional
  properties create a single point of failure?}

\htext{
The results of this paper already go a long way
to improving robustness of a single root-of-trust, compared to the state
of the art: lowering failure probability; mitigating and recovering
from the consequences of failure.  
}
The logical next step would be to
try and prevent failures in the first place.
However, the failure of a simplex system of reasonable complexity
cannot be prevented.
%As mentioned before, as future work, we plan to address this problem by ensuring dependability properties such as availability (e.g., through replication).   

\hdifftext{
Nevertheless, note that logical centralization is
not necessarily physical centralization.
And indeed, our plan for future work (and the way we drafted our architecture
paved the way) is to leverage state-of-the-art security and
dependability mechanisms using replication.  For instance, to achieve
tolerance of Byzantine faults, we can readily
enhance \ANCHOR by replication, taking advantage of state machine
replication libraries such as BFT-SMaRt~\cite{bessani2014bft},
replicating and diversifying components to prevent failure of this
logically central point, with the desired confidence.  
These concepts
have been applied to root-of-trust like configurations similar to
\ANCHOR~\cite{zhou2002csd,Cachin2004sdd,kreutz2014tsd}.
Furthermore, systems designed with state machine replication in mind
can also handle different types of threats, such as DoS, without 
compromising the operation of the
service~\cite{Kreutz2016acra}.
}

\hdifftext{
\textit{Won't the natural hardware evolution be by itself enough to reduce the penalty imposed by cryptographic primitives?} 
The recent reality seems to contradict this assertion --  hardware evolution does not seem enough, for several reasons.
First, new hardware architectures can benefit different existing software-based solutions.
For instance, both NaCl and OpenSSL take advantage of hardware-based AES accelerators.
Second, and as is well known, the fixed price of advancements in hardware seems to be coming to an end~\cite{ieee2015moore}.
This is made clear by most of the major IT companies, such as Google and Microsoft, to be redesigning existing software as a response to cope with this problem~\cite{Livshits2015DSM}. %,Verma2015LCM}.
%In short, hardware will not be the panacea.
%Simple, secure and robust software will (arguably) always lead to savings on CAPEX and OPEX.
}

\textit{Aren't traditional PKI and TLS implementations enough?}
Following what is becoming recurrently advocated by many in the industry and in the security community, we have tried to argue that the simplicity and size of software and IT infrastructure matters~\cite{cisco2014asr,Verizon2015dbir}. %kiravuo2013esec,
Higher complexity has been shown to lead inevitably to an increased likelihood of bugs and security incidents in software. 
Indeed, different implementations of PKI and TLS have been recently used as powerful ``weapons'' for cyber-attacks and cyber-espionage~\cite{pwc2014ucc,bocek2015iha}, leading to concerns about their robustness.
%~\cite{Meulen2013DigiNotar,Kaminsky2010pki,pwc2014ucc,bocek2015iha}.
Contrary to what this argument may suggest, that does not mean PKI and TLS are ``broken''.
We believe they remain fundamental to various IT infrastructures.
However, as the main challenges of securing SDN are usually relatively constrained to within a network domain, we have come to understand that simpler, domain-specific solutions seem to be preferable in this environment when compared to complex infrastructures such as the PKI, and large code bases as OpenSSL.

\textit{Wouldn't the use of out-of-band control channels solve most problems?}
Out-of-band channels may be useful in some contexts, but they are not ``intrinsically'' secure.
It is a recurrent mistake to consider physical isolation, \textit{per se}, as a form of security.
Several studies have indeed argued the opposite: that out-of-band channels worsen the problem, by making control plane management more complex and less flexible, endangering control plane communications~\cite{edwards2014researchers,Manousakis2015aap}.
%~\cite{swire2006theory,Hoepman2007IST,edwards2014researchers,Manousakis2015aap}
We do not take a stance in this discussion, but the fact is that real incidents, such as NSA sniffing of Google's cables between data centers~\cite{schneier2015data}, seem clear examples that out-of-band channels are not, \textit{per se}, enough.

\fi

\subsection{Other use cases of ANCHOR}
%%%%%%%%%%%%%%%%%%%%%%%%%%%

\iftrue
\hdifftext{
\textit{Using} \ANCHOR \textit{beyond control plane communications.}
As already alluded to in Section~\ref{sec:rwork}, \ANCHOR can be extended to support other use cases. 
For instance, one application running on top of the SDN controller could be required to provide proper credentials to identify itself.
Once successfully authenticated, it should have access to a specific set of system attributes defined by the operator during registration (e.g., read, write, notify, among other system calls~\cite{ferguson2013participatory,aliyu2017atrust}).
Towards this goal, different controllers could rely on authentication and authorization attributes globally enforced by \ANCHOR.
%During the registration process, the system administrator could define the set of attributes of each application.
Another interesting use case for \ANCHOR would be to offer security support for controller clustering. 
This is a timely problem.
To give an example, the current release of OpenDaylight does not provide encryption or authentication of control messages exchanged among controller instances~\cite{odl2018security}.
Since each controller instance would need to be registered with \ANCHOR, it would be possible to provide the same security mechanisms and services we grant to the southbound connection, to ensure security in east-west communication between controllers.
}

% FIXME: This work proposes a general framework of centralized management of non-functional properties of a SDN network. The main body of the work, however, is concerned with security properties. While it is great to develop treatment for concrete security properties, it is not clear what are the other significantly different "non-functional" properties, and how will the framework address them.
\hdifftext{
\textit{Addressing other non-functional properties of SDN.}
The design of \ANCHOR is generic enough to accommodate non-functional properties beyond security, such as dependability or quality of service.
With respect to the former, \ANCHOR could help in modularising the problem of replicated control.
Specifically, \ANCHOR could be responsible for coordination between controller replicas, for instance by guaranteeing a strongly consistent view of the network across all instances.
Similar to our security use case, the additional modularity of such design would allow a clean separation of concerns that could simplify the design of the various components.
Recent proposals~\cite{botelho2016design} have indeed started following a similar design choice.  
\ANCHOR could also provide trusted measurement services for ensuring a certain level of service even in the presence of malicious forwarding devices. 
For instance, once a malicious forwarding device were detected~\cite{chi2015how,kamisinski2015flowmon}, \ANCHOR could automatically remove it from the list of legitimate devices, forcing the disconnection of those devices by the controllers of the network, which would be registered to receive such events.
The subsequent topology updates on the controllers would trigger automatic traffic re-routing to ensure the quality of service of applications.
%As a second example, \ANCHOR could also host failure detection services to address data plane and control plane dependability issues.
%Those services can be used for triggering controller instance replacement or recovery after successful detection of a misbehaved or failed controller~\cite{zhang2017dealing}.
%After a failure, a controller instance should be recovered by automatic means if possible.
%\ANCHOR could provide such kind of services for SDN (e.g., remote control of virtual machines or host control through IPMI over LAN).
}

\fi

% Conclusion

%%%%%%%%%%%%%%%%%%%%%%%%%%%%%%%%%%%%%%%%%%%%%%%%
\section{Concluding remarks}
\label{sec:conclusion}
%%%%%%%%%%%%%%%%%%%%%%%%%%%%%%%%%%%%%%%%%%%%%%%%

In this paper, we debated the problem of enforcing
non-functional properties in SDN, such as security or dependability.
Re-iterating the successful philosophy behind the inception of SDN
itself, we advocate the concept of logical centralization of SDN
non-functional properties provision, which we materialize in terms of
the blueprint of an architectural framework, \ANCHOR.

Taking `security' as a proof-of-concept use case, we have
shown the effectiveness of our proposal. We made a gap analysis of
security in SDN and proposed solutions, by  populating the \ANCHOR middleware with
crucial mechanisms and services to fill those gaps and enhance the
security of SDN.

We evaluated the architecture, especially focusing on the
security-performance analysis tradeoff, giving proofs of the
algorithms, cryptographic robustness analyses, and experimental
performance evaluations. By resorting to lightweight yet secure primitives, we outperform the most
widely used encryption of OpenSSL by 50\%, with a higher level of
security.  Our solution also fulfills eleven of the security requirements
recommended by ONF.

The mechanisms we propose are certainly not the final answer to SDN
security problems.  That is not our claim.  \htext{ We however
  believe, and hope to have justified in the paper, that an
  architecture that logically centralizes non-functional
  properties of an SDN, has the potential to address some of the
  most preeminent unsolved problems regarding the robustness of the
  infrastructure.  We thus hope our work to trigger an important
  discussion on these fundamental architectural aspects of SDN.  }

%%%%%%% PJV: ack to fnr is already in page 1. Save this sec for acks
%%%%%%% of people, names or anonymous reviewers...
%%
%% \begin{acks} This work is partially supported by the University of
%%Luxembourg - SnT and by the Fonds National de la Recherche
%%Luxembourg (FNR) through PEARL grant FNR/P14/8149128. 
%%\end{acks}
%%

%\newpage

\bibliographystyle{ACM-Reference-Format}
\bibliography{refs_anchor}

\newpage

%Appendix
\appendix

%%%%%%%%%%%%%%%%%%%%%%%%%%%%%%%%%%%%%
%\section{Correctness of  Algorithm~\ref{alg:entropyupdate}}
\section{A source of strong entropy}
\label{appendix:entropyupdate}
%%%%%%%%%%%%%%%%%%%%%%%%%%%%%%%%%%%%%

{\color{black}

}

\textbf{Correctness.} We argue about the properties of
Algorithm~\ref{alg:entropyupdate}, as a source of strong entropy.

%%%%%%%%%%%%%%
\begin{lem}
\label{lem1_entropy}
If the initial values of rand\_bytes() and H(data) are
indistinguishable from random, then the resulting initial external
entropy (e\_entropy - line 2) is indistinguishable from random. Then,
the initial internal entropy (i\_entropy - line 3) will be also
indistinguishable from random.
\end{lem}

\begin{proofm}
 Assuming that rand\_bytes() uses one of the strongest pools of
 entropy of an operating system, such as /dev/urandom, the outcome of
 this function call will be indistinguishable from random.  Assuming
 that H is a cryptographically strong hashing function, the output of H(data) will be
 indistinguishable from random for every different input data.
 Consequently, the XOR operation between rand\_bytes() and H(data)
 will result in an indistinguishable-from-random initial e\_entropy.
 Following, the XOR operation between rand\_bytes() and e\_entropy
 will result in an indistinguishable-from-random initial i\_entropy.
 In other words, both internal and external entropy are initialized
 with indistinguishable-from-random values.
\end{proofm}

%%%%%%%%%%%%%%
\begin{lem}
\label{lem3_entropy}
If $P_i$, $P_j$, and i\_entropy are indistinguishable from random,
then the updated external entropy (e\_entropy - line 5) will be
indistinguishable from random.
\end{lem}

\begin{proofm}
As discussed before, the pools of entropy $P_i$ and $P_j$ contain
unpredictable events of external sources of entropy, such as network
traffic and idleness of links.  Thus, assuming that H is a
cryptographically strong hashing function, then the output of
H($P_i$||$P_j$) will be indistinguishable from random.
Lemma~\ref{lem1_entropy} shows that the internal entropy ($i\_entropy$)
is indistinguishable from random.  In consequence, the updated
external entropy ($e\_entropy$ - line 5), which is the output of an XOR
operation between two indistinguishable-from-random values, will be
indistinguishable from random.
\end{proofm}

%%%%%%%%%%%%%%
\begin{lem}
\label{lem2_entropy}
If the initial value of rand\_bytes() is indistinguishable from
random, then the resulting internal entropy (i\_entropy - line 7) is
indistinguishable from random.
\end{lem}

\begin{proofm}
The proof of Lemma~\ref{lem1_entropy} establishes that the output of
$rand\_bytes()$ is indistinguishable from random. Additionally,
$E\_counter$ is an internal counter not known by external entities.
Therefore, assuming that H is a cryptographically strong hashing
function, then $i\_entropy$ output by H($rand\_bytes() ||
E\_counter$) will be indistinguishable from random.
\end{proofm}

%%%%%%%%%%%%%%
\begin{theo}
\label{lem4_entropy}
If e\_entropy and i\_entropy are indistinguishable from random, then
the resulting entropy returned by entropy\_get (line 8) will be
indistinguishable from random.
\end{theo}

\begin{proofm}
Lemmata~\ref{lem1_entropy} and~\ref{lem3_entropy} show that the initial
and updated external entropy are indistinguishable from random.
Lemma~\ref{lem2_entropy} has shown that the internal entropy generated
in line 7 is indeed indistinguishable from random.  As a consequence,
$entropy$, as the output of an XOR operation between $i\_entropy$ and
$e\_entropy$ (line 8) will be indistinguishable from random.
This
proves that Algorithm~\ref{alg:entropyupdate}
satisfies property \textbf{Strong Entropy}.
\end{proofm}

%%%%%%%%%%%%%%%%%%%%%%%%%%%%%%%%%%%%%
%\section{Correctness of  Algorithm~\ref{alg:PRG}}
\section{Pseudorandom generator (PRG)}
\label{appendix:PRG}
%%%%%%%%%%%%%%%%%%%%%%%%%%%%%%%%%%%%%

{\color{red}

}

\textbf{Correctness.} We argue about the properties of
Algorithm~\ref{alg:PRG},  as a source of indistinguishable-from-random
pseudo-random values.

%%%%%%%%%%%%%%
\begin{lem}
\label{lem1_prg}
If entropy\_get() returns an indistinguishable-from-random value, then
the initial $seed$ (line 2), $counter$ (line 3) and pseudo random
value ($nprd$ - line 4) will be indistinguishable from random.
\end{lem}

\begin{proofm}
Theorem~\ref{lem4_entropy} establishes that the output of
$entropy\_get()$ is indistinguishable from random.  Thus, both the
$seed$ and the first $nprd$ will be indistinguishable from random.
Similarly, the function $long\_uint()$ (using as input
$entropy\_get()$ - line 3), which, on most architectures, uses 64 bits
to represent an unsigned long int, will return the value $counter$,
indistinguishable from random.
\end{proofm}

%%%%%%%%%%%%%%
\begin{lem}
\label{lem3_prg}
If entropy\_get() returns a value indistinguishable from random, then
the refreshed PRG internal state (lines 6-8) will lead to
indistinguishable from random values for $seed$, $counter$ and $nprd$.
\end{lem}

\begin{proofm}
The proof follows the same argumentation of the proof of
Lemma~\ref{lem1_prg}, for $seed$ and $counter$. As for $nprd$,
assuming that neither the seed or counter are known outside the PRG,
and assuming that H is a cryptographically strong hashing function,
then the output of H, having as input a concatenation of the new
$seed$, current $nprd$, and new $counter$, will be indistinguishable
from random.
\end{proofm}

%%%%%%%%%%%%%%
\begin{theo}
\label{lem2_prg}
If seed and nprd are indistinguishable-from-random values, then the
next nprd returned by PRG\_next (line 12) will be indistinguishable
from random.
\end{theo}

\begin{proofm}
Lemmata~\ref{lem1_prg} and~\ref{lem3_prg} established that both the
$seed$ and $nprd$ are always indistinguishable from random, since the
initial state.  Assuming that HMAC is a cryptographically strong
message authentication code primitive, and that the counter is not
known outside of the PRG, then the output of HMAC, keyed by $seed$ and
having as input a concatenation of $nprd$ and $counter$, will be
indistinguishable from random.
This
proves that Algorithm~\ref{alg:PRG}
satisfies property \textbf{Robust PRG}.
\end{proofm}

{\color{black}
%%%%%%%%%%%%%%%%%%%%%%%%%%%%%%%%%%%%%
\section{The three stages of ANCHOR}
\label{appendix:ANCHORstages}
%%%%%%%%%%%%%%%%%%%%%%%%%%%%%%%%%%%%%

Figure~\ref{fig:phases} illustrates the three stages of \ANCHOR, namely, setup, normal operation, and post-compromise recovery.
After setup and post-compromise recovery, it goes to normal operation.
The details of normal operation (e.g., device registration and association) are discussed in Sections~\ref{sec:devreg} and~\ref{sec:devassoc}. 
The complete post-compromise recovery protocol is presented in Section~\ref{sec:post_compromise}.

\begin{figure}[ht]
   \centering
   \includegraphics[width=0.90\columnwidth]{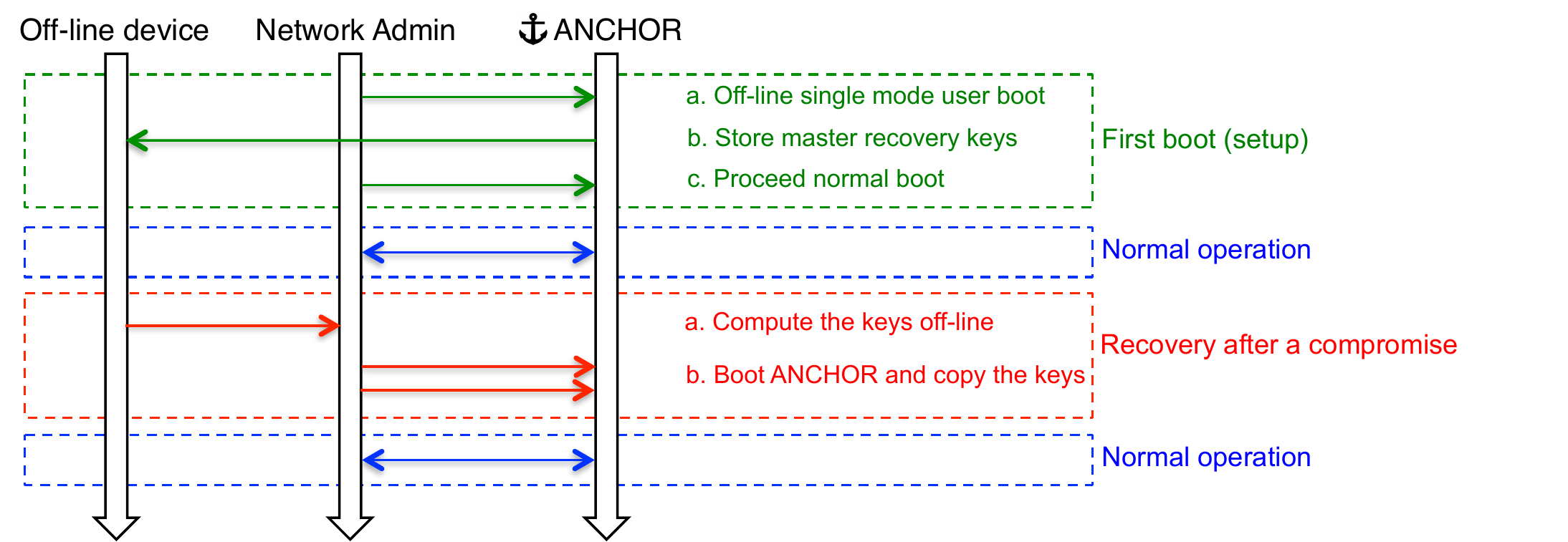}
   \caption{Setup, normal operation and PCR}
   \label{fig:phases}
\end{figure}

%%%%%%%%%%%%%%%%%%%
\subsection{Setup}
%%%%%%%%%%%%%%%%%%%

During the setup, three things happen:

\begin{enumerate}

\item \textit{Off-line single mode user boot}. The first boot should be off-line to generate the master recover keys safely. These keys need to be generated a single time and stored in a safe place.

\item \textit{Store master recovery keys}. The network admin should store the master recovery keys, for future use in case of a compromise, in an off-line device (e.g., USB stick). This device should be kept as secure as possible.

\item \textit{Normal boot}. After generating and safely storing the master recovery keys, the network admin can proceed the normal boot of \ANCHOR. This boot is going to bring up all services and functionalities of \ANCHOR and put it online, ready for use.

\end{enumerate}

%%%%%%%%%%%%%%%%%%%
\subsection{Normal operation}
%%%%%%%%%%%%%%%%%%%

The normal operation represents the phase in which \ANCHOR should be most of the time, i.e., online and fully operational.
The normal operation phase can happen after a first boot (setup phase) or after a recovery from a compromised state. 

%%%%%%%%%%%%%%%%%%%
\subsection{Recovery after a compromise}
%%%%%%%%%%%%%%%%%%%

To recover \ANCHOR after a compromise, the network admin has to:

\begin{enumerate}

\item \textit{Compute the keys off-line} using the master recovery keys. The network admin must recursively generate the network manager recovery keys and the device recovery keys. These are special purpose keys used to automatically and safely recover communications between \ANCHOR and all other entities, i.e., without needing additional procedures such as device re-registration. For more details on how it works, see Section~\ref{sec:post_compromise}.

\item \textit{Boot} \ANCHOR \textit{ and copy the keys}. After recursively computing the master recovery keys of managers and devices, the network admin should proceed a normal boot of the system and copy these keys into \ANCHOR.

\end{enumerate}

}

\iftrue

%%%%%%%%%%%%%%%%%%%%%%%%%%%%%%%%%%%%%
\section{Correctness of  Algorithm~\ref{alg:devAssoc}}
\label{appendix:devAssoc}
%%%%%%%%%%%%%%%%%%%%%%%%%%%%%%%%%%%%%

\textbf{Correctness.} We now formalize and prove the properties of Algorithm~\ref{alg:devAssoc}.
%%%%%%%%%%

As a result of the registration process, \ANCHOR keeps lists of
registered devices and controllers, and lists of the controllers each
device is authorized to associate with.

%%%%%%%%%%%%%%
\begin{prop} 
\label{prop1_algoAiD}
Any device F can only associate to a controller C authorized by the \ANCHOR.
\end{prop}

\begin{proofm}
Forwarding devices will be able to associate only to controllers
listed in the CList(F) provided by A (step 2 of
Algorithm~\ref{alg:devAssoc}), since if F tries to associate with a
non-authorized controller (for F), A will not proceed past step 4
after being contacted by that controller, aborting the association.
On the other hand, a rogue controller posing to F as authorised in
reply to
step 3, cannot jump to step 6 and invent an association key $AiD$ that
convinces F, since it does not know $x_f$.
This proves that Algorithm~\ref{alg:devAssoc} satisfies
property \textbf{Controller Authorization}.
\end{proofm}

%%%%%%%%%%%%%%
\begin{prop} 
\label{prop2_algoAiD}
Any device F can associate to some controller, only if F is authorized by the \ANCHOR.
\end{prop}

\begin{proofm}
Only if a device F is legitimate, i.e. it is in the list of registered
devices, will it be able to associate to some registered controller.
A will not proceed past step 1 of Algorithm~\ref{alg:devAssoc} after
being contacted by a rogue device, aborting the association.  On the
other hand, a rogue device posing to C as legitimate and authorised in
step 3, will make C proceed with step 4, indeed, but the request will
be rejected by A, since $E_F()$ is not recognisable by A,
corresponding to no shared key with a legitimate device. The replay
of an old (but legitimate) encrypted $E_F()$ request in step 3 will
also fail, since it is bound to the (current) nonces.
This proves that Algorithm~\ref{alg:devAssoc} satisfies
property \textbf{Device Authorization}.
\end{proofm}

%%%%%%%%%%%%%%
\begin{prop} 
\label{prop3_algoAiD}
At the end of Algorithm~\ref{alg:devAssoc} execution, the association
ID ($AiD$) is only known to F and C.
\end{prop}

\begin{proofm}
A creates $AiD$ in step 5, and forgets about it after sending it to C (see Section~\ref{sec:devassoc}).
% FIXME: explain this in the text.
$AiD$ is sent from A to C, encrypted both by $Ke_{AF}$ and $Ke_{AC}$, keys
shared by A only with F and C respectively.  C trusts it came from A,
due to the HMAC, so the two encrypted blocks should contain the same
$AiD$ value, and sends the $AiD$ under $Ke_{AF}$ encryption to F.
So, at the end of the execution of the
algorithm, both F and C, and only them, hold $AiD$.
This proves that Algorithm~\ref{alg:devAssoc} satisfies
property \textbf{Association ID Secrecy}.
\end{proofm}

%%%%%%%%%%%%%%
\begin{prop} 
\label{prop4_algoAiD}
At the end of Algorithm~\ref{alg:devAssoc} execution, the seed
($SEED$) is only known to F and C.
\end{prop}

\begin{proofm}
C creates $SEED$ in step 7.  $SEED$ is sent from C to F, encrypted by
$K_AiD$, association key known only to C and F, as per
Proposition~\ref{prop3_algoAiD}.  C trusts that F, and only F, has the
same $SEED$ sent, when it receives back from F the XOR of $SEED$ with
the current nonce $x_g$ encrypted with $AiD$, since (as per
Proposition~\ref{prop3_algoAiD}) only F could have opened the
encryption of $SEED$ with $AiD$ in the first place, and encrypt the
reply.
This proves that Algorithm~\ref{alg:devAssoc} satisfies
property \textbf{Seed Secrecy}.
\end{proofm}

\fi

%%%%%%%%%%%%%%%%%%%%%%%%%%%%%%%%%%%%%
\section{ONF's security requirements}
\label{appendix:secReq}
%%%%%%%%%%%%%%%%%%%%%%%%%%%%%%%%%%%%%

Several security requirements should be fulfilled in control
plane communications.  Most of these requirements are enumerated in
ONF's best practice recommendations~\cite{onf2015pps}.  In this appendix we go
through the eleven (out of twenty four) such requirements that are addressed by the \ANCHOR, iDVV and NaCl.

\textit{Both communicating devices should be authenticated (REQ 4.1.1)}. 
Using our \ANCHOR, all devices have to be properly registered and authenticated before proceeding any other operation.

\textit{Operations (e.g., association) of components should be authorized (REQ 4.1.2)}. 
The \ANCHOR needs to explicitly authorize associations between any two devices.
Each association has a unique identification.

\textit{Devices should agree upon the security (e.g., key materials) associations (REQ 4.1.3)}.
By using the \ANCHOR and its mechanisms, such as the source of strong entropy, we ensure strong key materials.
The iDVV mechanism is initialized by the two communicating devices once the association has been authorized by the \ANCHOR.

\textit{Integrity of packets should be ensured (REQ 4.1.4)}.
We provide integrity and authenticity of packets through message authentication codes. 
By default, we generate one iDVV per packet, providing stronger security.

\textit{Each device should have a unique ID and other devices should be able to verify the identity (REQ 4.2.1)}.
Devices are uniquely identified by the \ANCHOR. 
%Forwarding devices are uniquely identified by their DPID, while controllers receive a unique ID from the \ANCHOR.
The unique IDs are associated to the devices as soon as they are registered within the \ANCHOR.

\textit{Issues related to the lifecycle of IDs should be managed, such as generation, distribution, maintenance, and revocation (REQ 4.2.2)}.
The \ANCHOR provides the services required for managing device IDs.
IDs are assigned to devices during the registration phase.
Revocation can be done by network administrators at any time.

\textit{Devices should be able to verify the integrity of each message (REQ 4.4.4)}.
Any two communicating devices are able to verify the integrity of each message through message authentication codes. % and the iDVV linked to the respective association.

\textit{Amplification effects should be taken into account, i.e., attackers should not be able to perform reflection attacks (REQ 4.4.5)}.
We use requests and replies of the same size between devices and the \ANCHOR, which avoids reflection attacks.

\textit{Automated key/credential management should be implemented by default, allowing generation, distribution, and revocation of security credentials (REQ 4.8.3)}.
We have in place automated mechanisms for refreshing credentials, such as refresh the iDVV's seed using the \ANCHOR's source of strong entropy.

%Identification, authentication and authorization of devices , we ensure through our \ANCHOR.
%Every device has to be registered within the \ANCHOR before being allow to communicate with other devices.
%All device-to-device associations have to be authorized by the \ANCHOR, i.e., faked devices are not allowed to communicate with other devices, for instance.
%Communications between devices and the \ANCHOR and between two devices are authenticated using the iDVV mechanism.

\textit{Data confidentiality, integrity, freshness and authenticity} are ensured by the integrated device verification value.
iDVVs are used to encrypt data and generate message authentication codes.
Additionally, iDVVs can also be used as nonces, ensuring data freshness.

\textit{Availability} is ensured by recommending multiple controllers to the forwarding devices.
This is one of the essential tasks of the \ANCHOR.

Lastly, it is also worth mentioning that whilst we do not meet all
security requirements of ONF's guidelines, we do meet the fundamental
ones with regard to security. For instance, requirements such as REQ
4.4.2, REQ 4.4.3, REQ 4.7.1, REQ 4.7.2, and REQ
4.7.3~\cite{onf2015pps} are not yet covered by our architecture and
protocols.  However, most of these requirements are related to rate
control of messages, additional signaling messages for dealing with
future network attack types, and accountability and traceability.
Such kind of requirements can be added (in the future) without
impairing our conceptual architecture.  In fact, some of these
requirements, such as rate control of messages, are technical, rather
than conceptual, which can be addressed with the right
amount of engineering.

\end{document}